\documentclass[amsmath,amssymb,prd,superscriptaddress,a4paper,nofootinbib,11pt]{revtex4}
\usepackage{amsmath}
\usepackage{amssymb}
\usepackage{graphicx,pstricks}
\usepackage[mathscr]{eucal}
\usepackage{color}
\usepackage{slashed}
\usepackage{xspace}
\usepackage{bm}
\usepackage{ulem}
\usepackage{cancel}
\usepackage{enumerate}
\usepackage{longtable}
\usepackage{units}
\usepackage{hepnames,hepparticles}
\usepackage{hyperref}
\usepackage{cleveref}
\usepackage{array}

\newcommand{\spa}[1]{\ensuremath{\widetilde{#1}}}

\newcommand{\GeV}{\text{GeV}}

\newcommand{\RpV}{\text{RPV}\xspace}
\newcommand{\de}{\mathrm{d}}

\newcommand{\CheckMATE}{{\tt CheckMATE}\xspace}
\newcommand{\ETmis}{E_T^{\text{miss}}}
\newcommand{\abs}[1]{\left| #1 \right|}

\newcommand{\sigmastable}{\sigma^{\mathrm{95\%}}_{\mathrm{stable}}}
\newcommand{\sigmactau}{\sigma^{\mathrm{95\%}}_{c\tau}}
\newcommand{\mmed}{m_{\mathrm{mediator}}}
\newcommand{\mLL}{m_{\mathrm{LL}}}
\newskip\LTleft \LTleft=0cm
\newskip\LTright \LTright=0cm
\newdimen\LTcapwidth \LTcapwidth=\textwidth
\newcommand{\AddrBonn}{%
Bethe Center for Theoretical Physics \& Physikalisches Institut der 
Universit\"at Bonn, \\
 53115 Bonn, Germany
}

\newcommand{\AddrSoton}{%
School of Physics \& Astronomy, University of S., Southampton SO17 1BJ, United Kingdom
}

\newcommand{\AddrSTFC}{%
Rutherford Appleton Laboratory, 
Science \& Technology Facilities Council (STFC),
Chilton, Didcot. Oxon OX11 0QX, United Kingdom
}

\graphicspath{{./Figs/}}

\begin{document}

\hfill BONN--TH--2015--14, XXXX--XXXX \vspace{0.2cm}

\title{Hunting for neutral, long-lived exotica at the LHC
using a missing transverse energy signature.}

\author{Alexander Belyaev}
 \email{a.belyaev@soton.ac.uk}
 \affiliation{\AddrSoton}

\author{Stefano Moretti}
 \email{s.moretti@soton.ac.uk}
 \affiliation{\AddrSoton}

\author{Kilian Nickel} 
\email{nickel@th.physik.uni--bonn.de}
\affiliation{\AddrBonn}
 
 \author{Marc C.~Thomas}
 \email{m.c.thomas@soton.ac.uk}
 \affiliation{\AddrSoton}

 \author{Ian Tomalin}
 \email{ian.tomalin@stfc.ac.uk}
 \affiliation{\AddrSTFC}


\begin{abstract}
Searches at the Large Hadron Collider (LHC) for neutral, long-lived particles have historically relied on the
detection of displaced particles produced by their decay {\it within} the detector volume. 
In this paper  we study the potential of the complementary  signature comprising of the missing transverse energy ($\ETmis$) signal, traditionally used to look for dark matter, e.g., the lightest supersymmetric particle (LSP), to extend the LHC coverage to models with long-lived (LL) particles when they decay {\it outside} the detector.
Using CMS and ATLAS analyses  at the  8 TeV LHC, 
we set an upper limit at the 95\% confidence level (CL) on the production cross sections for  two specific scenarios:
(i)  a  model with a heavy non-standard model Higgs boson decaying to a LL scalar and (ii)
an R-parity violating (\RpV) SUSY model with a LL neutralino. 
 We show that this method can significantly  extend the LHC sensitivity to neutral, LL particles with arbitrary large  lifetimes and that the limits obtained from a $\ETmis$ signal are comparable to those from displaced particle searches for decay distances above a few meters. 
Results obtained in this study do not not depend on the specific decay channel of the LL particle
and therefore are model-independent in this sense.
We provide  limits for the whole two-dimensional plane in terms of the mass of the LL particle and 
the mass of the mediator up to masses of 2 TeV including particular benchmarks studied in the original experimental papers.
We have made these limits 
available in the form of a grid which can be used for the interpretation
of various other new physics models.
\end{abstract}
\maketitle

\section{Introduction}

New long-lived (LL) particles are predicted 
by various Beyond Standard Model (BSM) scenarios, such as
neutralinos in Supersymmetry (SUSY) with weak $R$-parity violation \cite{Barbier:2004ez}, gluinos in split-SUSY \cite{Hewett:2004nw},
``hidden valley'' models \cite{Han:2007ae}, 
heavy photons in Little Higgs  models with T-parity~\cite{Cheng:2004yc,Cheng:2003ju}
broken by a Wess-Zumino-Witten anomaly term~\cite{Hill:2007zv}
and LL heavy neutrinos in the minimal $B-L$ extension of the Standard Model (SM) \cite{Basso:2008iv}.

In this paper, we focus on the case of neutral LL particles. Searches for neutral LL particles at the LHC have historically been reliant on reconstructing 
their decay products within the detector volume. If the LL particle lifetime is of order picoseconds to nanoseconds, then its decay can yield 
striking signatures of displaced leptons, jets, photons or charged tracks. Numerous searches for these
signatures have been performed at the LHC, exemplified by \cite{Khachatryan:2014mea,CMS:2014hka,CMS:2014wda}
(CMS) and \cite{Aad:2015rba, Aad:2015asa} (ATLAS).
However, for longer lifetimes, an increasing proportion of the LL particles decay outside the detector, reducing
the efficiency of these searches and leading to correspondingly weaker cross section limits. 

In this paper, we extend existing limits to arbitrarily long lifetimes, by exploiting the fact that neutral, LL particles that decay outside the 
detector will be visible as missing transverse energy, $\ETmis$. As such, our approach is complementary to the traditional one exploiting
displaced particles. In fact, the cross section limits obtained using the $\ETmis$ signature will improve with increasing lifetime, as a larger 
proportion of particles decay outside the detector. 

To illustrate this method, we concentrate on the results of two CMS papers, which searched for displaced vertices reconstructed within the CMS tracker, produced by either two leptons \cite{CMS:2014hka} or a 
quark-antiquark pair \cite{CMS:2014wda}. 
In both these papers, limits were set for a number of benchmark points for two specific models: (i) a simplified model with a heavy, non-SM Higgs boson $H^0$ decaying into two LL scalar bosons $X$ which then decay exclusively to either $e^+ e^-$, $\mu^+ \mu^-$ or $q \bar q$ and (ii) a SUSY model with a LL neutralino $\tilde{\chi}^0$ decaying via a $R$-parity violating coupling exclusively to either $\ell^+ \ell^- \nu$ or $q \bar{q}^{(\prime)} \mu$ (with the prime indicating different quark flavours). For both of these models, we use measurements of the $\ETmis$ signature
from CMS and ATLAS analyses at 8 TeV, to set upper bounds on signal cross sections for each decay channel, assuming  that the LL particle is stable. Using the geometric properties of the detectors and the energy and rapidity distribution of the LL particle, we then extrapolate these cross section limits to finite lifetimes including when the mean decay distance is within the detector. For each benchmark point in these CMS papers, we have extended the limits such that an upper limit on the cross section is provided for any lifetime from around $10^{-2}$ cm (depending on  benchmark point) up to arbitrarily long lifetimes. We show that for these models, analysis of $\ETmis$ signature can set more stringent cross section limits than displaced vertex searches for a LL particle with a lifetime of order a few nanoseconds and longer. Upper limits over a range of masses of the LL particle and its mediator are also provided for both models, under the assumption that the LL particle has a lifetime such that it always decays outside the detector.

The rest of the paper is organised as follows. In Sec.~\ref{sec:setup} we discuss the models, signal simulation and 
details of the analyses.
The results are given in Sec.~\ref{sec:results}, followed by our conclusions in Sec.~\ref{sec:conclusion}.

\section{Setup}
\label{sec:setup}

\subsection{Models}
\label{sec:models}

To best demonstrate how $\ETmis$ signatures can be used to extend to longer particle lifetimes, the cross 
section limits obtained from LHC displaced particle searches, we use the same signal models that were 
studied by CMS in \cite{CMS:2014hka, CMS:2014wda}. Both these CMS papers considered the same pair of 
signal models, examining LL particle decays to leptonic  \cite{CMS:2014hka} or hadronic \cite{CMS:2014wda} 
final states, respectively.

It is important to stress whereas in the CMS simulations, the LL particles were allowed to decay, for our 
$\ETmis$ study, we instead use simulations in which they are defined as completely stable. This is because 
we (conservatively) assume that neutral LL particles only contribute to the $\ETmis$ signature if they both leave 
the detector before decaying. As a result, which final state they eventually decay to is completely 
irrelevant to the $\ETmis$ study. This makes our analysis much more model-independent than traditional 
displaced particle searches at the LHC.
Further, our analysis does not depend on the reconstruction or identification efficiencies
of the LL particle decay products, which allows us to have good sensitivity to the signature under study.

The two signal models are as follows.

\begin{figure}[htpb]
  \centering
  \includegraphics[width=0.9\linewidth]{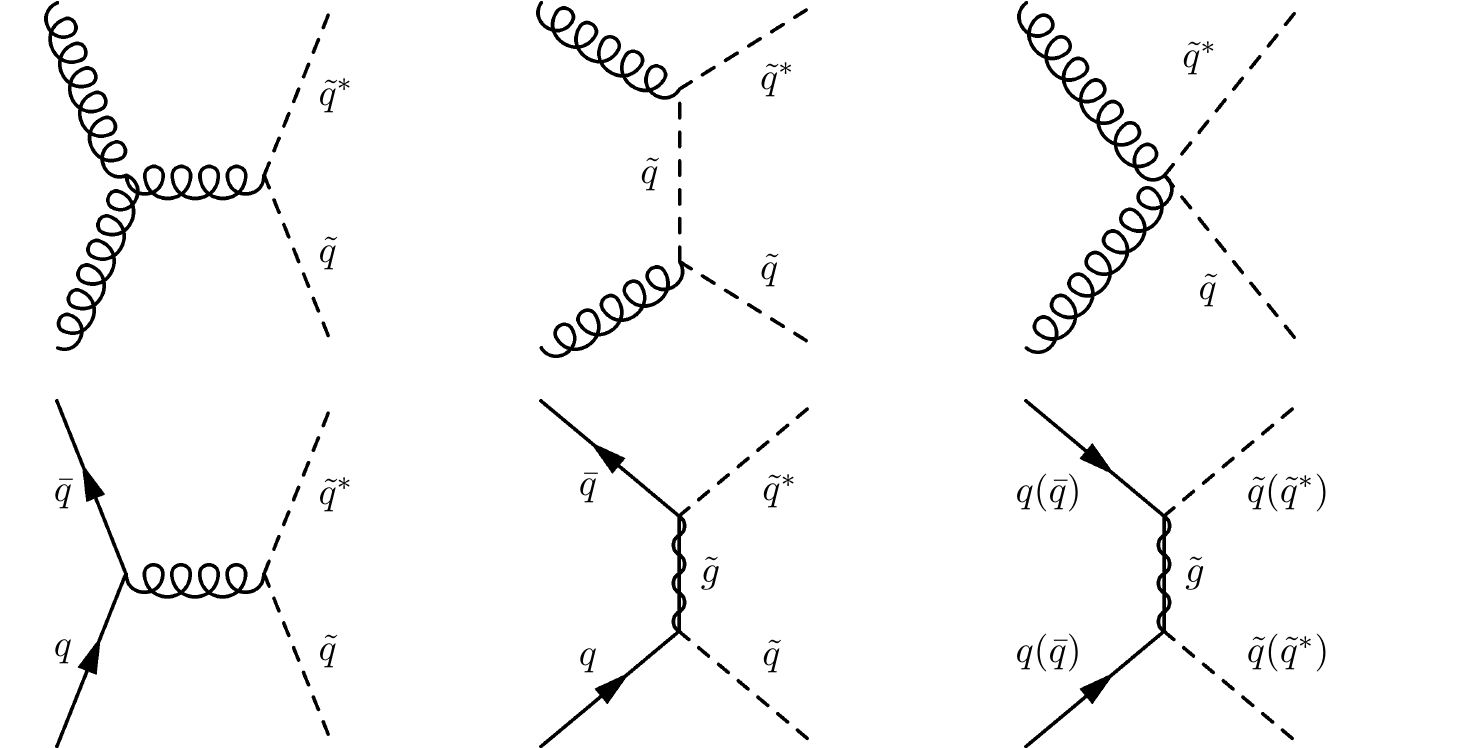}
  \caption{Strong production of squark pairs in SUSY.}
  \label{fig:SquarkProd}
\end{figure}

\begin{itemize}
\item[(1)] A simplified model with a heavy, non-SM Higgs boson $H^0$ produced via gluon fusion (via an effective vertex from $\tfrac{1}{2}\mathrm{Tr}[G^2]H$, with $H$ being a new singlet) and decaying to two long-lived, heavy, neutral, spinless bosons $X$. In the CMS analyses, these subsequently decay either to two leptons \cite{CMS:2014hka} or a quark-antiquark pair \cite{CMS:2014wda}, whereas our simulation treats them as stable,
as explained above.
\begin{align}
  g g &\to H \to X X\\
  X &\to e^+e^-, \mu^+ \mu^-, q\bar{q} {\textrm{~~~(in the CMS analyses)}}
\end{align}
The respective  production diagrams are shown in Fig.~\ref{fig:Hprod}. The decay width of the heavy Higgs is assumed to be  much smaller than its mass, $\Gamma_H \ll m_H$ and so we use the narrow width approximation. Thus we only consider processes where the heavy Higgs is produced on-shell and the mass relation $m_X \leq \tfrac{1}{2} m_H$ holds. We will refer to this model as HXX for short. 
\begin{figure}[htpb]
  \centering
  \includegraphics[width=0.9\linewidth]{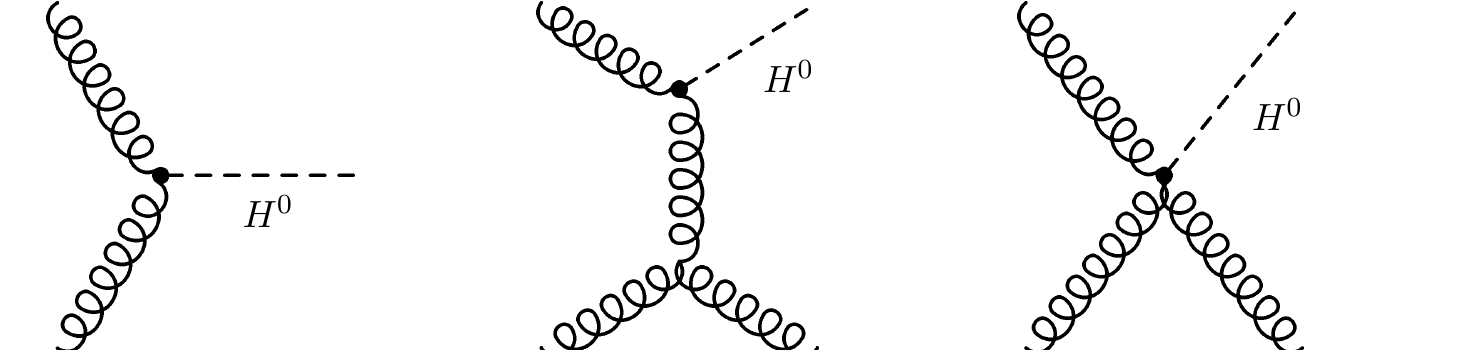}
  \caption{Production of $H^0$ in the HXX model, including up to 1 jet in the final state.}
  \label{fig:Hprod}
\end{figure}
\item[(2)] A SUSY model with small $R$-parity violating (\RpV) couplings which give rise to a LL neutralino $\tilde\chi^0$. A pair of squarks $\tilde q$ of arbitrary flavour ($\tilde u_L$) is strongly produced and decays into a quark $q$ and a neutralino $\tilde \chi^0$,
\begin{equation}
  p p \to \tilde{q} {\tilde{q}^*}, ~~~~ \tilde{q} \to q \tilde{\chi}^0.
\end{equation}
In the CMS analyses, the neutralino decays either to $\ell^+ \ell^- \nu$ \cite{CMS:2014hka} or to $u \bar{d} \mu^-$ \cite{CMS:2014wda}, via $\lambda_{ijk}  \hat L_i\hat L_j\hat E_k$ or $\lambda^{\prime}_{ijk}  \hat L_i\hat Q_j\hat D^c_k$ $R$-parity violating terms, respectively \cite{Barbier:2004ez}. Our simulation instead
treats the neutralino as stable, as explained above.
 The diagrams for the strong production of the squark pair are shown in Fig.~\ref{fig:SquarkProd}. There are three types of squark pairs: $\tilde q\tilde q$, $\tilde q\tilde q^*$ or $\tilde q^*\tilde q^*$. We denote these squark pairs as $\tilde Q\tilde Q$ with $\tilde Q=\tilde q,\tilde q^*$. The branching ratio of $\tilde q \to q \tilde \chi^0$ is assumed 1 for all events. 
\end{itemize}

\subsection{Event generation}
\label{sec:mcevent}
As we have already mentioned, we look for signals where the neutral LL particle ($X$ or $\tilde{\chi}^0$) leaves the detector before decaying.
For the HXX model the only way to observe the $\ETmis$ signal is to use events with a high $P_T$ jet  from initial state radiation (ISR)
induced by strong interactions. In this case this high $P_T$ jet will be recoiling against the pair of $XX$ neutral LL particles
providing a classic mono-jet signature for $XX$ decaying outside of detector.
In case of the \RpV SUSY scenario, if the squark and neutralino  have a small mass gap, one  can again rely on a
mono-jet signature. In contrast, if the mass gap is large, then the squark will decay to a SM  quark and a neutralino,
which would lead to a `$\ETmis$ + jets' signature.
The signal events for both of the models in \ref{sec:models} were generated by {\tt MadGraph5 v2.1.2} \cite{Alwall:2014hca} with {\tt Pythia 6.4} \cite{Sjostrand:2006za,Torrielli:2010aw} for parton showering and hadronisation. 

For the $\RpV$--SUSY model, we use the default {\tt mssm} model from {\tt MadGraph} (because we only consider signatures where the neutralino decays outside the detector) and used $\tilde q=\tilde u_L$ without loss of generality to generate the strong production of squark pairs $\spa q \spa q$, $\spa q^*\spa q^*$ and  $\spa q\spa q^*$ (which we denote as $\tilde Q\tilde Q$) decaying via $\spa q \to q \tilde{\chi}_1^0$ ($\spa q^* \to \bar q \tilde{\chi}_1^0$).
In order to ensure accurate simulation of hard ISR jets, we allow an additional matrix element jet in our event generation which is matched using the $k_T$ MLM scheme \cite{Torrielli:2010aw}. All SUSY masses (except $m_{\tilde\chi^0},m_{\tilde q}$) including the gluino mass are set to 5~TeV to match the model used by CMS.

The HXX scenario  described in sec.~\ref{sec:models} was implemented using  a  model 
generated using the LanHEP  package \cite{Semenov:1996es,Semenov:2002jw,Semenov:2010qt}, with an effective vertex between gluons and the heavy Higgs implemented via a $\tfrac{1}{2}\text{Tr}[G^2]H$ Lagrangian term, where $G$ is the usual gluon field strength tensor. Again $k_T$ MLM matching between 0-jet and 1-jet events was used to ensure accurate simulation of the hard jets.

\subsection{Used CMS and ATLAS $\ETmis$ Analyses}
\label{sec:analyses}
Signatures with  $\ETmis$ have been the focus of the dedicated papers by ATLAS and CMS, mainly is the context of SUSY searches involving the LSP.
These papers present results for various final states produced in association with the $\ETmis$. Which paper allows us to place the strongest 
cross section limits on the two signal models we consider will depend both on the model and on the particle masses. 
We therefore need to implement and use multiple such publications to ensure that we place the tightest bounds possible over the range of masses in our models.
Fortunately, the results of a large number of such papers have already been implemented in the  {\tt CheckMATE} \cite{Drees:2013wra,deFavereau:2013fsa,Cacciari:2011ma,Cacciari:2005hq,Cacciari:2008gp,Read:2002hq,Lester:1999tx,Barr:2003rg,Cheng:2008hk} framework,  which allows easy use of the implemented searches. This tool takes a given sample of Monte Carlo events in the HEP or HEPMC format after PYTHIA/HERWIG level of simulation
and  performs a detector simulation for the sample using {\tt Delphes-3 \cite{deFavereau:2013fsa}}.
{\tt CheckMATE} is then able to apply in turn each analysis as described in the experimental papers to the generated signal event. The resulting efficiencies along with information provided by the publication, such as the 95\% C.L. on signal count $S^{95}_{\text{exp}}$, is used to produce results from which we can find the cross-section limit placed on our model by each analysis. Those analyses which we used have all been validated by using published results including available cut-flows. 

For each paper, CMS and ATLAS typically give results for a number of different signal regions, for example, defined by different $\ETmis$ 
requirements. We will refer to these as `analysis sets'. When we place limits on one of our signal models for given particle masses, we only
use the results of the best analysis set within the best paper, where `best' is defined as the one yielding the strongest {\it expected limit}.
By using the expected limit instead of the observed one, we avoid the `look elsewhere' effect. In producing our limits, we do not
account for any systematic uncertainty on the new physics signal selection efficiency, as this would be model dependent.

From the long list of available papers, three are particularly important,  setting the best limits for the models we studied. 
A very brief outline of the selection cuts and analysis sets of these three is given below. (A fourth paper, a monojet analysis by CMS
\cite{Khachatryan:2014rra}, is potentially interesting, but not yet available inside \CheckMATE).

\begin{enumerate}
\item ATLAS $\ETmis$ + multi-jet analysis \cite{TheATLAScollaboration:2013fha}. \\
It uses 20.3 fb${}^{-1}$ of $\sqrt s = 8$~TeV data. $\ETmis$ must be above $\unit[160]{\GeV}$, the leading jet must have $p_T(j_1)>\unit[130]{\GeV}$ and the second leading jet $p_T(j_2)>\unit[60]{\GeV}$. The analysis sets are distinguished  by jet multiplicity 2,3,4,5,6, corresponding to analysis set codes {\it A,B,C,D,E}, while only jets with $p_T>\unit[60]{\GeV}$ are valid in this count. Given one of these five categories, signals are then subjected to {\it loose (L), medium (M)} or {\it tight (T)} constraints. In our case, analysis sets {\it AM, BM, BT, CM, CT} are relevant. For full details, cf. {\bf Page 3, Table 1 of \cite{TheATLAScollaboration:2013fha}}.
\item ATLAS $\ETmis$ + monojet analysis \cite{Aad:2015zva}. \\
It uses 20.3 fb${}^{-1}$ of $\sqrt s = 8$~TeV data. Events must have at least one jet with $p_T>\unit[120]{\GeV}$ and $\abs{\eta}<2.0$ and no charged  leptons (of $p_T>\unit[7]{\GeV}$). For the leading jet, $p_T/\ETmis > 0.5$ must hold ($\ETmis > \unit[150]{\GeV}$ required). The number of jets is unrestricted, but the leading jet is only considered (monojet-like selection). Nine analysis sets are defined between $\unit[150]{\GeV}<\ETmis<\unit[700]{\GeV}$, labelled {\it SR1} through {\it SR9}. Complete definitions, cf. {\bf Page 7, Table 2, of \cite{Aad:2015zva}}
\item CMS analysis using the $\alpha_T$ variable \cite{Chatrchyan:2013mys}. \\
It uses 11.7 fb${}^{-1}$ of $\sqrt s = 8$~TeV data.
Instead of $\ETmis$, this analysis uses the related variable $\alpha_T$ \cite{Randall:2008rw,Khachatryan:2011tk} to suppress multijet background events. This variable is used to be more independent of mismeasurements of $\ETmis$. For two back-to-back jets with $E_T^{j_1}=E_T^{j_2}$, $\alpha_T$ is equal to 0.5. A value greater than 0.5 signifies that the jets are recoiling against significant $\ETmis$. For further details of the $\alpha_T$ variable see \cite{Chatrchyan:2013mys,Randall:2008rw,Khachatryan:2011tk}. Events with $e$ or $\mu$ with $p_T>\unit[10]{\GeV}$ are vetoed as well as those  with an isolated photon with $p_T<\unit[25]{\GeV}$. To cut out multijet background events, $\alpha_T>0.55$ is required. Also, the scalar sum of all transverse jet energies, $H_T=\sum_{i=1}^{n_{\text{jet}}}E_T^{j_i}$, must be larger than $\unit[275]{\GeV}$. The two leading jets must each have $p_T>\unit[100]{\GeV}$ and the leading jet satisfies $\abs{\eta}<2.5$, but these conditions are also relaxed for some analysis sets. 
The analysis sets are named after the number of jets ({\tt 23j\_} for 2-3 jets or {\tt 4j\_} for $\ge 4$ jets) + number of $b$-jets ({\tt 0b\_} or {\tt 1b\_}) + lower limit of $H_T$ bin ({\tt 275, 325, 375, 475} etc.). Example: {\tt 23j\_0b\_325}.

\end{enumerate}

\subsection{Escape Probability}
\label{sec:escapeprob}

In order to find the production cross section limits for long-lived particle pairs with a given lifetime, we calculate the proportion of events passing analysis cuts where both particles decay outside the detector, producing a missing transverse energy signature, and use this along with the limits if the same particles were stable, calculated as discussed in Sec.~\ref{sec:analyses}, to find the 95\% cross section limits as a function of the lifetime.

In order to achieve this, we edited the {\tt CheckMATE} code, so that for each simulated event which passed all of the selection cuts, we calculate for each LL particle the probability of it leaving the detector before decaying.
This probability is
\begin{equation}
p(D) = \exp \left( \frac{-D}{c\beta \gamma \tau}\right)
\end{equation}
where $D$ is the distance from the interaction point to the periphery of the detector, calculated using the size and shape of the detector and the flight direction of the LL particle, and $\beta$, $\gamma$, and $\tau$ are the usual relativistic factors and the lifetime of the particle. For this calculation, the ATLAS and
CMS detectors are assumed to be cylindrical in shape, with ATLAS having a length of 46~m and a diameter of
25~m, and CMS having a length of 21~m and a diameter of 15~m.
The event is subsequently weighted according to probability that both LL particles leaving the detector undecayed,
\begin{equation}
  w = p_1(D_1)p_2(D_2).
\end{equation}
with $p_1$, $p_2$ denoting the probabilities for particle 1 and 2 respectively.
Summing these weights allows us to calculate the proportion, $P$, of these events which would have given an $\ETmis$ signature. We are thus able to calculate the 95\% C.L. on the signal cross section for any arbitrary lifetime,
\begin{equation}
  \sigmactau = \frac{1}{P}\sigmastable
\end{equation}
where $\sigmastable$ is the cross section limits calculated using {\tt CheckMATE} by assuming that all the LL particles decay outside the detector as described in Sec.~\ref{sec:analyses}.
There is a simpler approximation to obtain the lifetime dependent limit $\sigmactau$, which is suitable for other researchers who  wish to quickly approximate similar limits as those presented in Figs.~\ref{fig:FFX_limit}-\ref{fig:RPV_limit}. This method requires only an energy distribution of the LL particles and is described in detail in the appendix \ref{sec:escapeprobApp}. We found this method to give a reasonably agreement with our more accurate results. It can be applied to the limits provided by our grid results.
N.B. Events where only one LL particle exits the detector before decaying are also likely to yield a missing
transverse energy signature, and could therefore be used to improve the limits obtained with this 
signature, particularly for smaller particle lifetimes. However, this has not been done here, because to do 
so would require an understanding of how the ATLAS and CMS detectors would react to the other LL particle: the
one decaying within the detector volume. Whether the decay products of this particle contribute to the
visible energy in the event depends on details of their event reconstruction algorithms and on the selection
requirements of their missing transverse energy publications.

\section{Results}
\label{sec:results}

We have performed analyses using the $\ETmis$  signature
for  the benchmark points (BPs) used in the CMS studies \cite{CMS:2014hka} and \cite{CMS:2014wda} 
of the displaced vertices, where the BPs correspond to various particle masses in the two signal models.
As a result we have  obtained the cross section limits $\sigmastable$, i.e., the cross section for $c\tau\to\infty$ for both models for the $\sigma(pp\to H^0\to XX)$ and $\sigma(pp\to \spa Q\spa Q \to  \tilde{\chi}^0  \tilde{\chi}^0+\mathrm{jets})$  processes,  respectively.
\begin{table}
\begin{tabular}{|c|c|c|c|c|} \hline
Benchmark Point & $m_{H}$ (GeV) & $m_{X}$ (GeV) & $\sigmastable$ (pb) & Analysis -- SR\\ \hline
1a & 125 & 20 & 57.8 & {\tt ATLAS monojet \cite{Aad:2015zva} - SR4}\\ \hline
1b & 125 & 50 & 39.9 &{\tt ATLAS monojet \cite{Aad:2015zva} - SR3} \\ \hline
2a & 200 & 20 & 17.9 & {\tt ATLAS monojet \cite{Aad:2015zva} - SR4}\\ \hline
2b & 200 & 50 & 21.3 & {\tt ATLAS monojet \cite{Aad:2015zva} - SR3}\\ \hline
3a & 400 & 20 & 3.55 & {\tt ATLAS monojet \cite{Aad:2015zva} - SR6}\\ \hline
3b & 400 & 50 & 6.03 & {\tt ATLAS monojet \cite{Aad:2015zva} - SR4}\\ \hline
3c & 400 & 150 & 3.81 & {\tt ATLAS monojet \cite{Aad:2015zva} - SR5}\\ \hline
4a & 1000 & 20 & 0.97 & {\tt ATLAS monojet \cite{Aad:2015zva} - SR6}\\ \hline
4b & 1000 & 50 & 0.71 & {\tt ATLAS monojet \cite{Aad:2015zva} - SR6}\\ \hline
4c & 1000 & 150 & 0.95 & {\tt ATLAS monojet \cite{Aad:2015zva} - SR7}\\ \hline
4d & 1000 & 350 & 0.80 & {\tt ATLAS monojet \cite{Aad:2015zva} - SR6}\\ \hline
\end{tabular}
\caption{Benchmark points from \cite{CMS:2014hka} and \cite{CMS:2014wda} (Model 1, HXX) and their 95\% CL upper limit on cross section, together with the CMS or ATLAS $\ETmis$ paper from which this limit was derived. \label{tab:benchmarkHXX}}
\end{table}
\begin{table}
\begin{tabular}{|c|c|c|c|c|} \hline
Benchmark Point & $m_{\tilde q}$ (GeV) & $m_{\tilde{\chi}^0}$ (GeV) & $\sigmastable$ (pb) & Analysis -- SR\\ \hline
1 & 120 & 48 & 33.5 & {\tt CMS $\alpha_T$ \cite{Chatrchyan:2013mys} - 4j\_0b\_325}\\ \hline
2 & 350 & 148 & 0.57  & {\tt CMS $\alpha_T$ \cite{Chatrchyan:2013mys} - 23j\_0b\_325}\\ \hline
3 & 700 & 150 & 0.041 & {\tt ATLAS multijet \cite{TheATLAScollaboration:2013fha} - AM}\\ \hline
4 & 700 & 500 & 0.24 & {\tt CMS $\alpha_T$ \cite{Chatrchyan:2013mys} - 23j\_0b\_375}\\ \hline
5 & 1000 & 148 & 0.0086 & {\tt ATLAS multijet \cite{TheATLAScollaboration:2013fha} - AM}\\ \hline
6 & 1000 & 500 & 0.025 & {\tt ATLAS multijet \cite{TheATLAScollaboration:2013fha} - AM}\\ \hline
7 & 1500 & 150 & 0.0018 & {\tt ATLAS multijet \cite{TheATLAScollaboration:2013fha} - CT}\\ \hline
8 & 1500 & 494 & 0.0024 & {\tt ATLAS multijet \cite{TheATLAScollaboration:2013fha} - CT}\\ \hline

\end{tabular}
\caption{Benchmark points from \cite{CMS:2014hka} and \cite{CMS:2014wda} (Model 2, $\RpV$--SUSY model) and their 95\% CL upper limit on cross section, together with the CMS or ATLAS $\ETmis$ paper from which this limit was derived. \label{tab:benchmarkRPV}
}.
\end{table}
The results for the HXX model are shown in Tab.~\ref{tab:benchmarkHXX} and for the $\RpV$-SUSY model in Tab.~\ref{tab:benchmarkRPV}, where we indicate the analysis set which provides the best sensitivity for each point. For the HXX model, for every BP, the ATLAS monojet + $\ETmis$ paper \cite{Aad:2015zva} 
provides the best sensitivity.
 This can be understood from the fact that the heavy Higgs is produced on-shell from gluon fusion and then decays to two back-to-back $X$ bosons (in the heavy Higgs rest frame), resulting in very little $\ETmis$ unless the heavy Higgs is boosted as a result recoiling against a jet from initial state radiation (ISR). 
For the \RpV-SUSY model, the paper providing the best limit depends on the BP. For $m_{\tilde{q}} = 120,350$, the CMS paper \cite{Chatrchyan:2013mys}, which uses the $\alpha_T$ variable, provides the best limit. On the other hand, for $m_{\tilde{q}} = 1000,1500$ GeV, the best limit is provided by using the ATLAS paper 
\cite{TheATLAScollaboration:2013fha}, which studies a large $\ETmis$ + multi-jet signal. Since in this
model, squarks are produced which then each decay to a quark and a LL particle, the presence of $\ETmis$ is 
not dependent on ISR in this case, so papers allowing for multiple jets in association 
with $\ETmis$ provide the best limits. The obtained for the HXX model are significantly weaker than those
obtained for the \RpV-SUSY model, because only a small fraction of events contain the hard ISR on which
the HXX limits rely. 

\begin{figure}[htbp]
\vskip -0.7cm
\makebox[\textwidth][c]{\hspace*{-0.5cm}\includegraphics[width=0.55\textwidth]{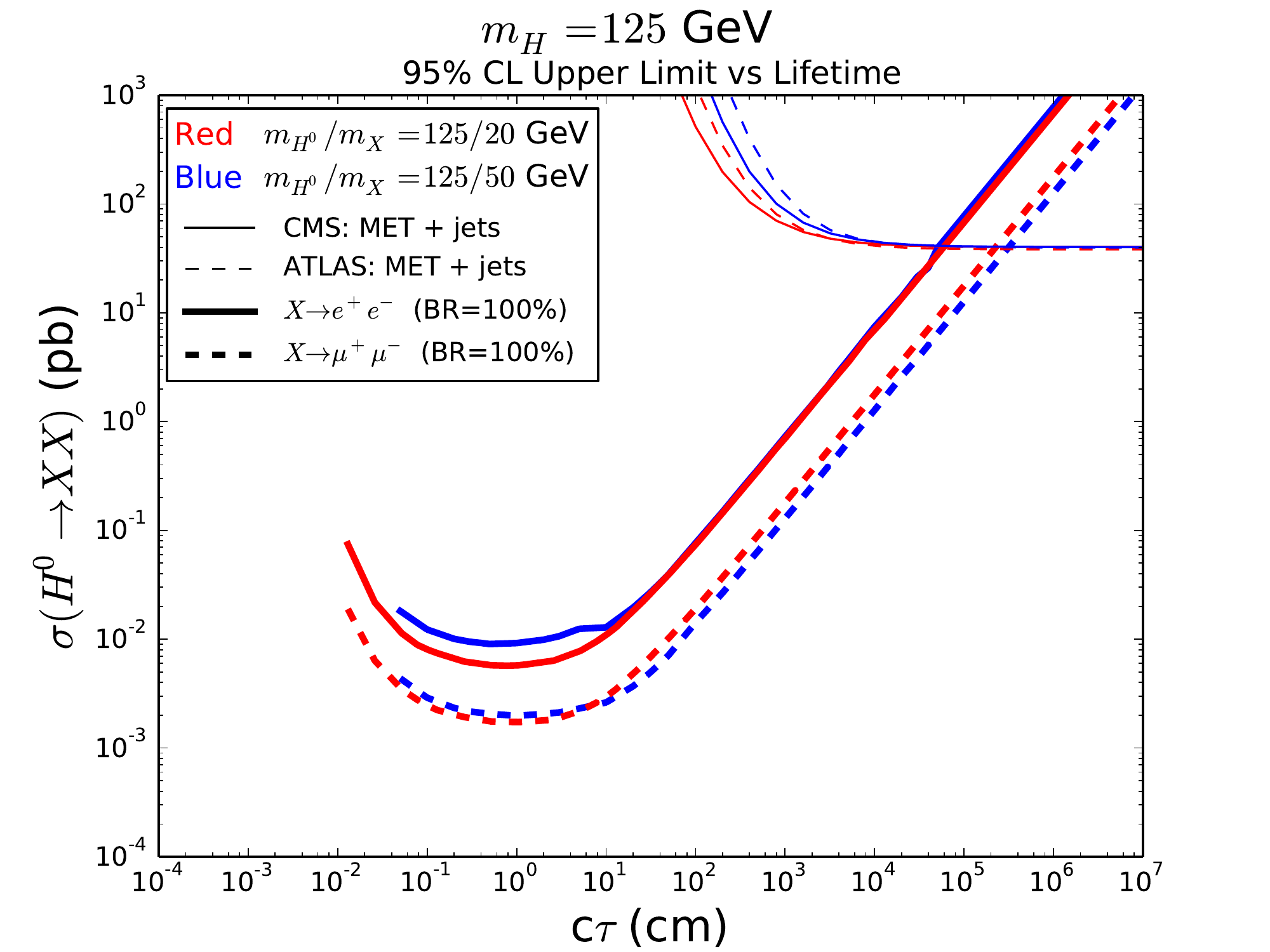}%
\hspace*{-0.5cm}\includegraphics[width=0.55\textwidth]{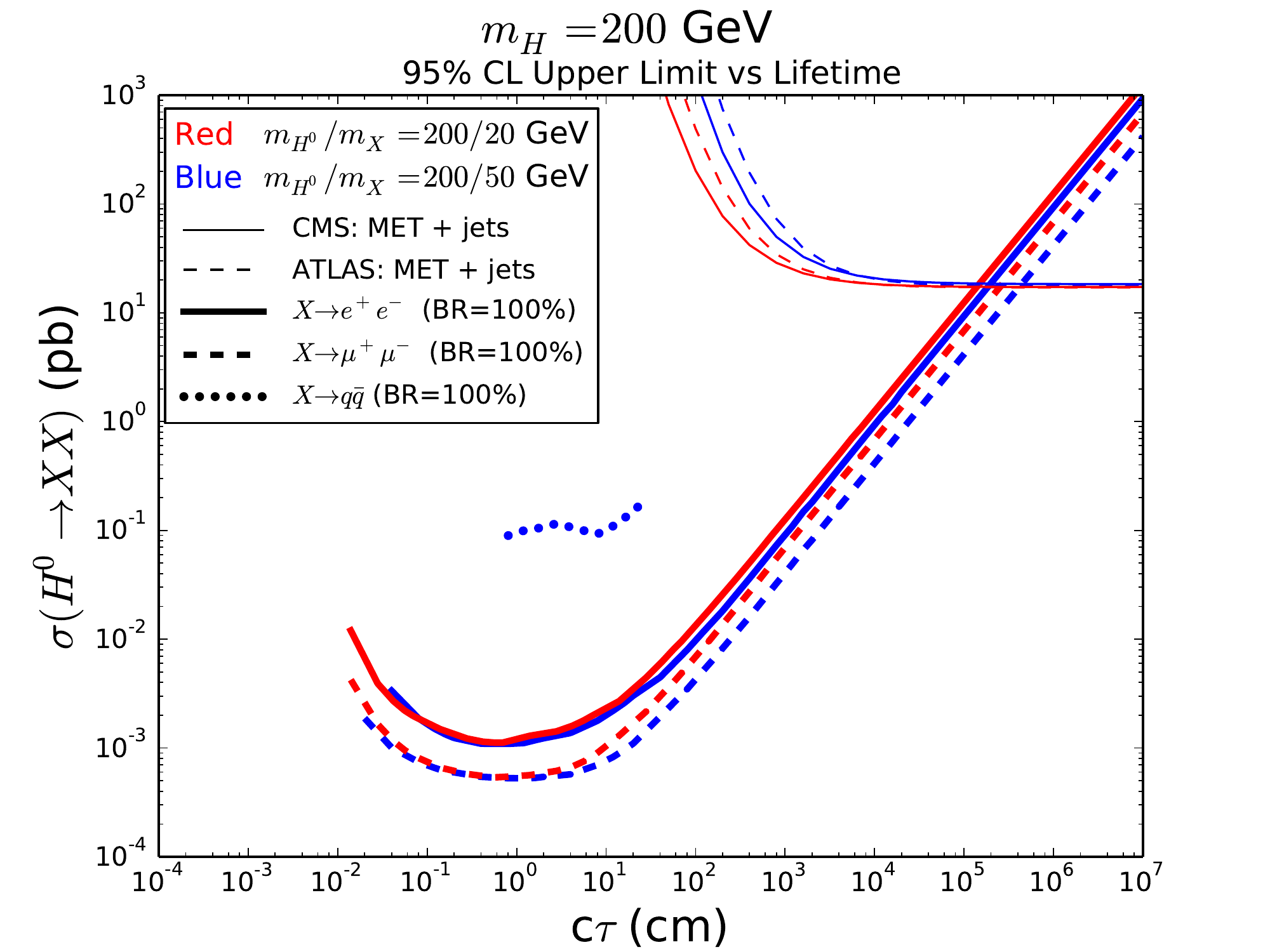}}
\\
\vskip -0.7cm
\hspace*{3cm}(a)\hspace*{0.5\textwidth}(b)
\\
\makebox[\textwidth][c]{\hspace*{-0.5cm}\includegraphics[width=0.55\textwidth]{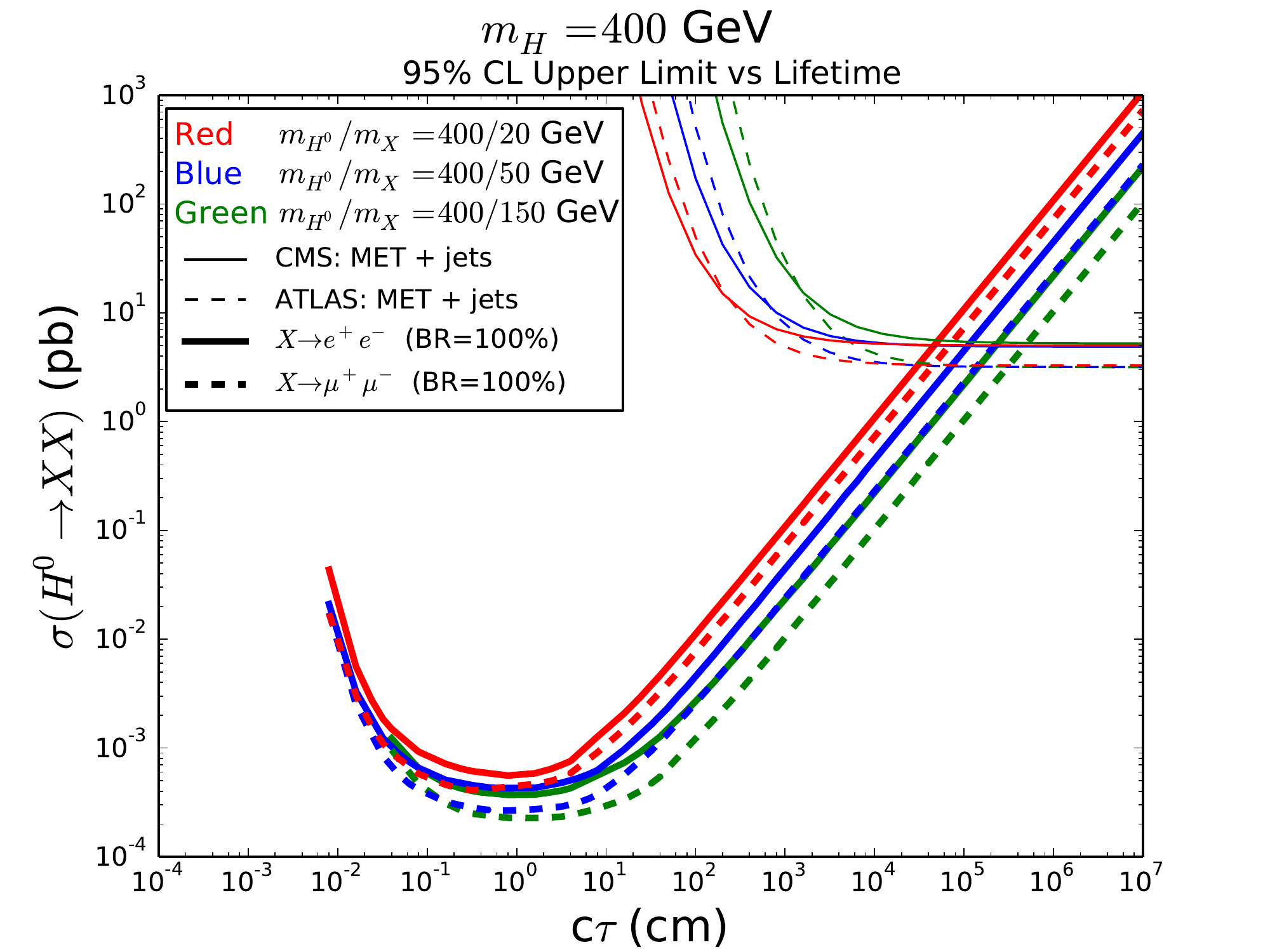}%
\hspace*{-0.5cm}\includegraphics[width=0.55\textwidth]{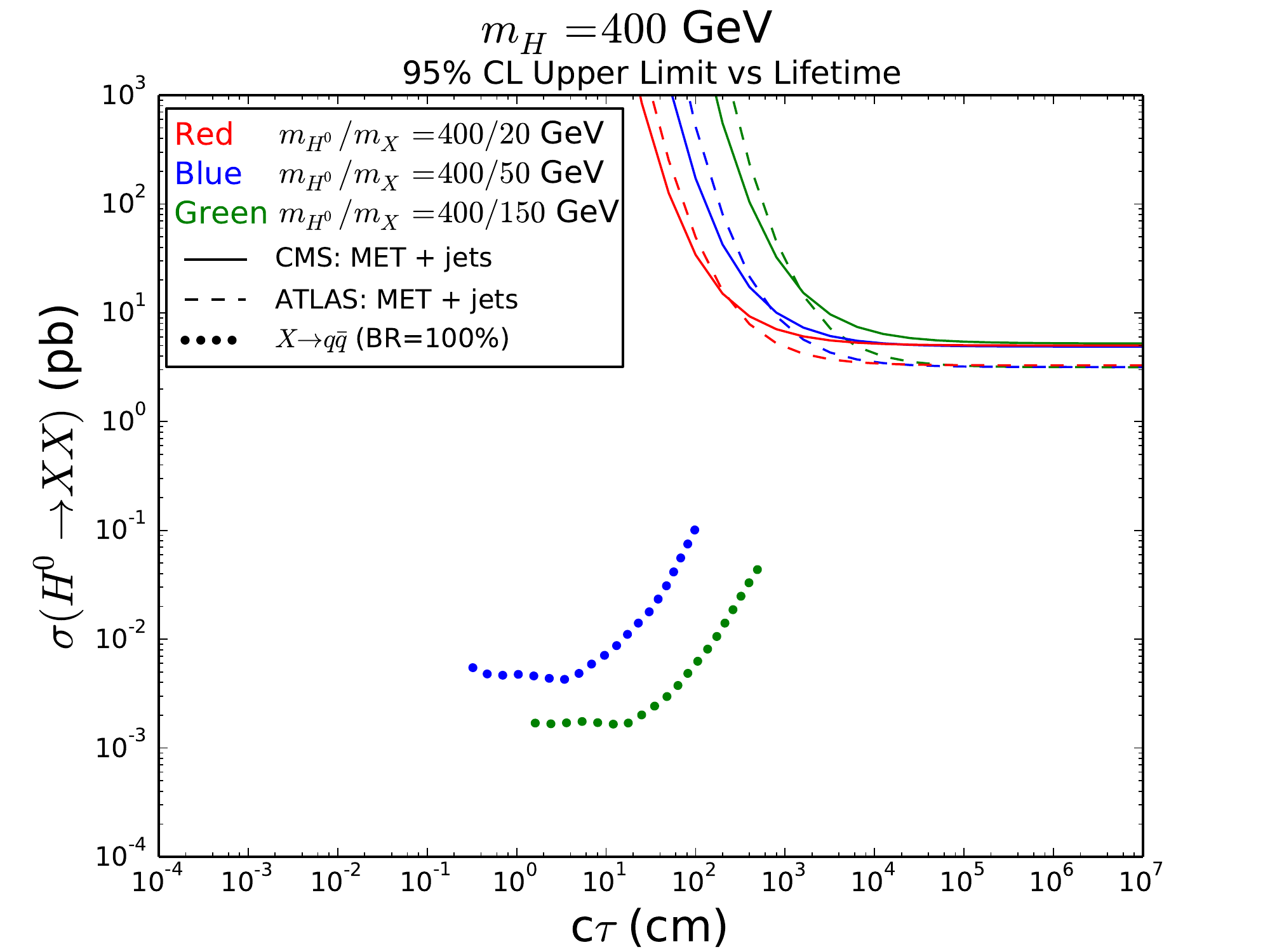}}
\\
\vskip -0.7cm
\hspace*{3cm}(c)\hspace*{0.5\textwidth}(d)
\\
\makebox[\textwidth][c]{\hspace*{-0.5cm}\includegraphics[width=0.55\linewidth]{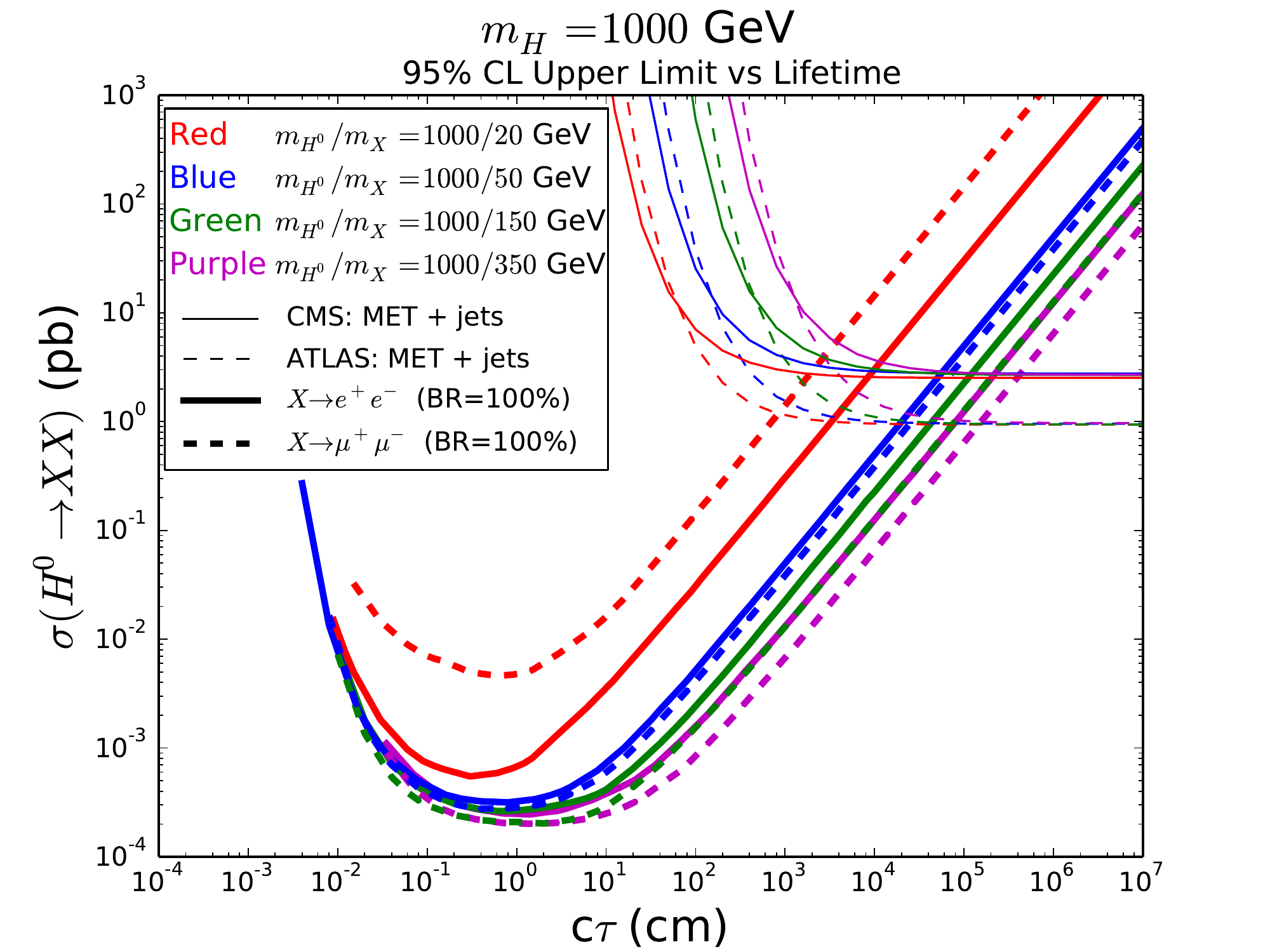}%
\hspace*{-0.5cm}\includegraphics[width=0.55\linewidth]{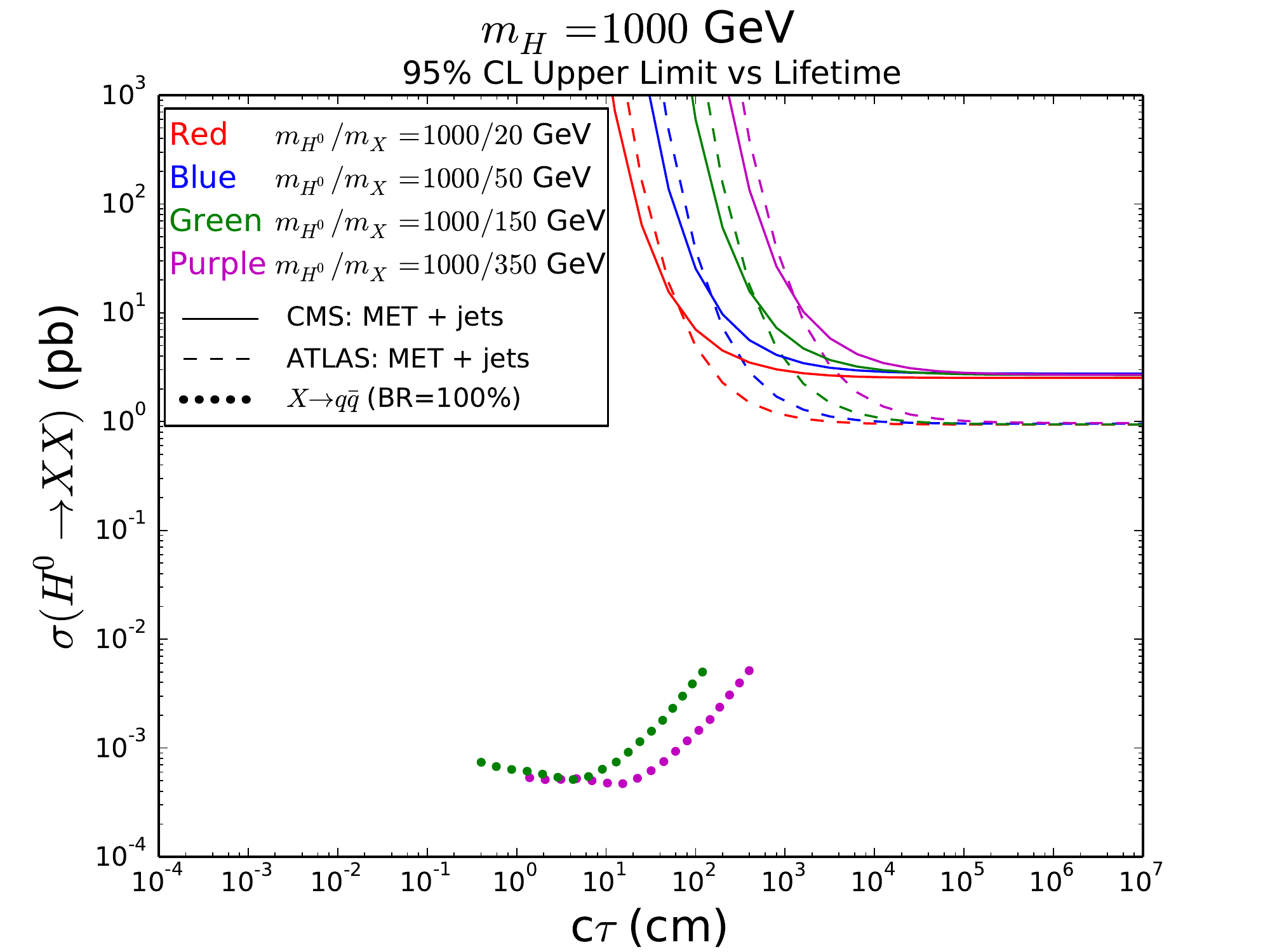}}
\\
\vskip -0.7cm
\hspace*{3cm}(e)\hspace*{0.5\textwidth}(f)
\vskip -0.4cm
\caption{95\% CL upper limits on cross sections for the heavy Higgs model (HXX) with $m_H = 125$ GeV (a), 200 GeV (b),
400 GeV (c,d) and 1000 GeV (e,f). The colour {\color{red} red} ({\color{blue} blue}) indicates $m_X=20$ GeV (50 GeV) {\it for all curves}. The thin curves in the upper-right corner of all 
figures show our new $\ETmis$-derived limits on LL particle  cross sections for each detector (solid: CMS, dashed: ATLAS). For comparison, the  cross section limits from the CMS displaced vertex searches, under the assumption of 100\% branching ratios, are shown by thick curves: displaced leptons searches ($X\to\ell^+\ell^-$) \cite{CMS:2014hka} are indicated by the solid curves for $\ell=e$ and by dashed curves for $\ell=\mu$; whereas 
displaced jet searches ($X\to q \bar q$) \cite{CMS:2014wda} are indicated by dotted curves.
Our new limits are identical in (c) and (d) as well as in (e) and (f) and have been split for clarity.
\label{fig:FFX_limit}}  
\end{figure}

\begin{figure}[htpb]
  \centering
 \makebox[\textwidth][c]{
  \includegraphics[width=0.6\linewidth]{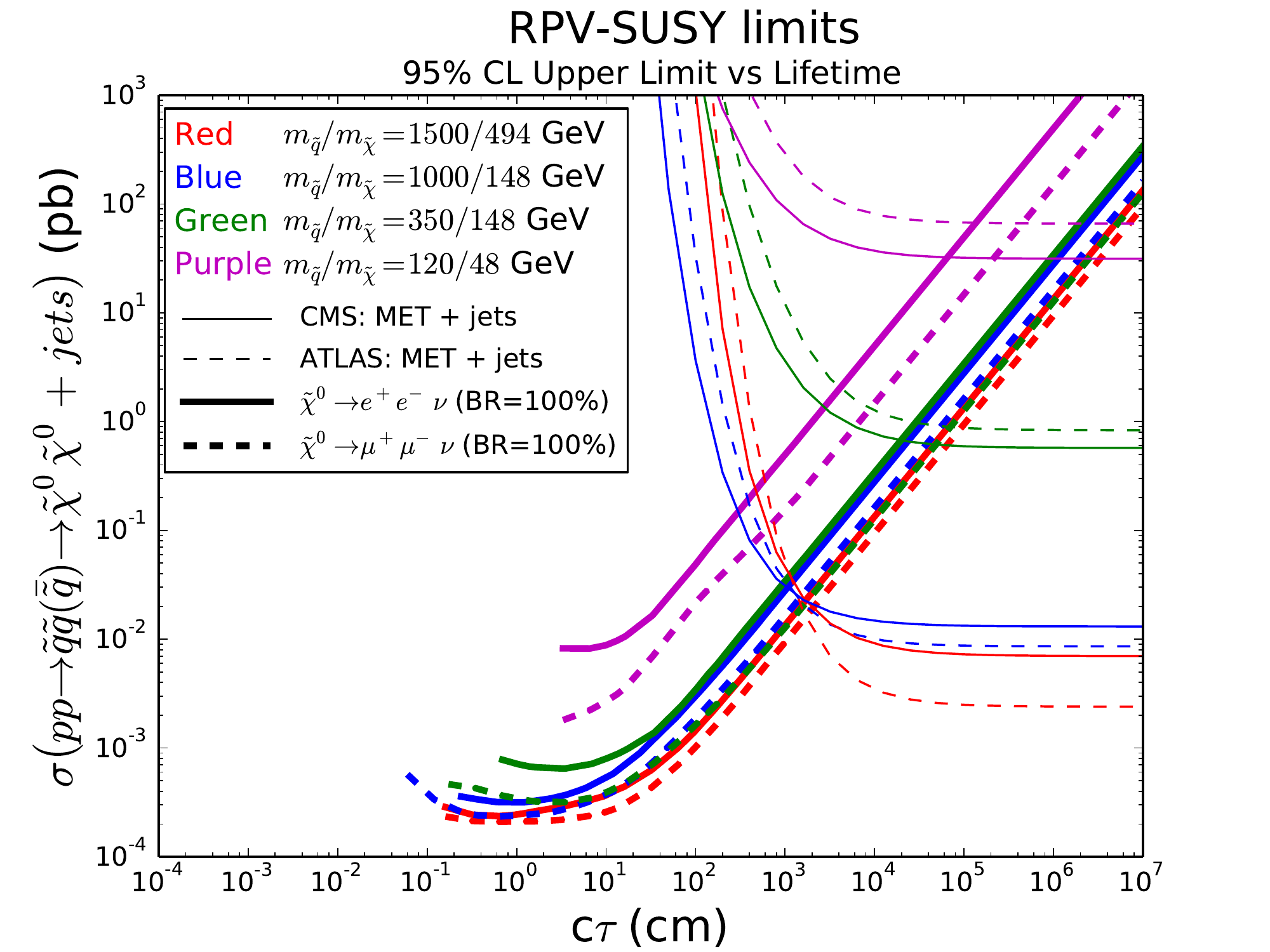}
  \hspace{-5mm}
  \includegraphics[width=0.6\linewidth]{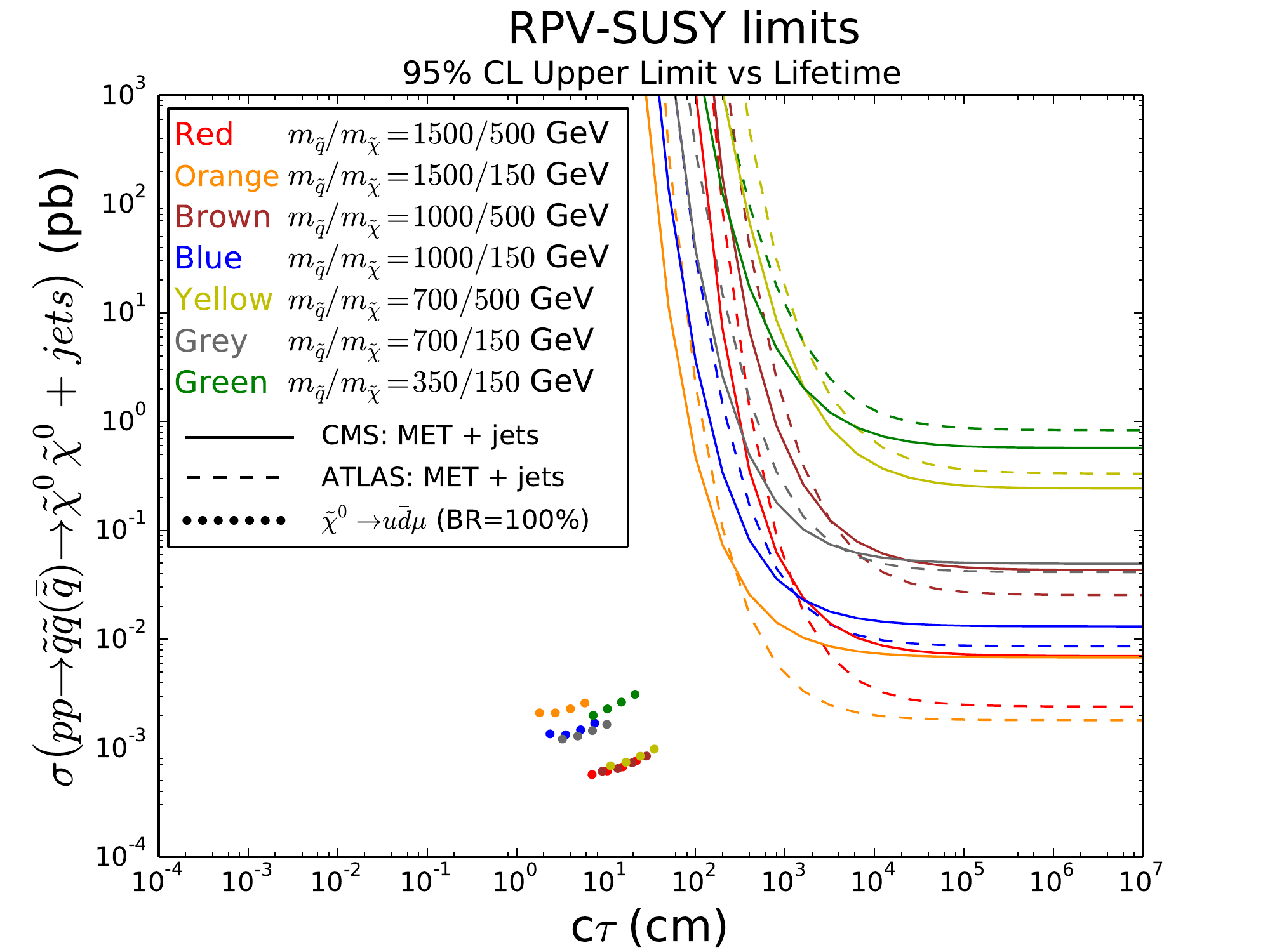}}
\hspace*{3mm}(a)\hspace*{0.56\textwidth}(b)
  \caption{95\% CL upper limits on cross sections for the $\RpV$--SUSY model with colours indicating various
mass points. The thin curves in the upper-right corner of both (a) and (b) show our new $\ETmis$-derived limits on LL particle  cross sections for each detector (solid: CMS, dashed: ATLAS). For comparison, the  cross section limits from the CMS displaced vertex searches, under the assumption of 100\% branching ratios are shown by thick curves: (a) for displaced dilepton searches ($\tilde{\chi}^0 \to \ell^+ \ell^- \nu$) \cite{CMS:2014hka}, with the solid curves indicating $\ell=e$ and the dashed curves indicating $\ell=\mu$; and (b) for displaced dijet searches \cite{CMS:2014wda} searches ($\tilde{\chi}^0 \to u \bar{d} \mu$) shown by dotted curves.
\label{fig:RPV_limit}}
\end{figure}

We subsequently calculate the upper 95\% CL on cross sections, $\sigmactau$, for arbitrary lifetimes by performing the procedure described in Sec.~\ref{sec:escapeprob}.
These results are plotted in Figs.~\ref{fig:FFX_limit},\ref{fig:RPV_limit}, where Fig.~\ref{fig:FFX_limit}
 show results for the HXX model for BPs with a heavy Higgs mass, $m_H$ of 125 GeV, 200 GeV, 400 GeV and 1000 GeV, and Fig.~\ref{fig:RPV_limit} displays the results for the \RpV-SUSY model. Each colour corresponds to a different BP, with the thin solid curves denoting the limits found using the dimensions of the CMS detector and using the CMS analyses, and the dashed thin line corresponding to the equivalent ATLAS limits. For comparison, we also plot the published results from the CMS displaced vertex analyses \cite{CMS:2014hka,CMS:2014wda} in thick curves (either solid, dashed or dotted). Beyond a certain lifetime, the cross section limits for the displaced vertex signatures increase in proportion to a power of the LL particle lifetime, and so appear on the log-log plot as a straight line. This can be understood from the following consideration.
Using a simplified  picture and assuming that the detector only has non-zero acceptance for particles decaying within a distance $L$ from the centre of the detector, the probability that a LL particle of momentum $P$ and mass $M$ decays within this acceptance region is $1 - \exp(\frac{-L M}{P c \tau})$, which tends to 
$\frac{-L M}{P c \tau}$ in the long lifetime limit. Cross section limits will scale in
inverse proportion to the acceptance. An analysis such as the CMS displaced dilepton vertex search 
\cite{CMS:2014hka}, which relies on the reconstruction of the decay products of
just one LL particle per event will thus yield cross section limits that scale in proportion to $\tau$ in
the large $\tau$ limit. This consideration allowed us to extrapolate these CMS limits to longer lifetimes than in their original publication, providing that the original results reached long enough lifetimes for this 
scaling behaviour to be observed. We did not attempt this for the CMS displaced dijet vertex search
\cite{CMS:2014wda}, because the original publication did not reach sufficiently long enough lifetimes in
that case.

The results show that, although the minimum cross section limits (occurring at $c \tau =$ \mbox{$\mathcal{O}$(1 cm)}) from displaced vertex searches are of order a fb, and those from the $\ETmis$ searches are of order a pb or more, the sensitivity to LL particles from the $\ETmis$ signature
can be  better than from displaced vertex searches for 
comparatively large times starting from 
$c \tau$ about  $10^3$ cm (for the \RpV-SUSY model with decays to $e^+ e^- \nu$ - see Fig.~\ref{fig:RPV_limit}, BP $m_{\tilde{q}} / m_{\tilde{\chi}^0} = 120/48$ GeV). However, generally the limits become comparable for $c \tau$ of $\mathcal{O}$($10^4 - 10^5$ cm).

When making such comparisons, it is important to note that in Figs.~\ref{fig:FFX_limit} and \ref{fig:RPV_limit}, the limits presented for the CMS displaced vertex searches are assuming a branching ratio of 1 of the LL particle to its respective decay, whilst our limit using $\ETmis$ searches is independent of decay channels or branching ratios. As an example, this means that in a realistic scenario where the LL $X$-particle has a branching ratio of 0.01 to $e^+ e^-$, the presented CMS displaced vertex limits would be weakened by 2 orders of magnitude, and the limits from $\ETmis$ signals would therefore be comparable for proper decay lengths as low a 1 metre or less for certain benchmark points. This highlights the fact that the limits set using $\ETmis$ are less model dependent than those for displaced vertices, thus they represent a new and complementary tool of investigation.

One should note about the results in Figs.~\ref{fig:FFX_limit} and  \ref{fig:RPV_limit} that, even when ATLAS provide the best result for $\sigmastable$, due to the smaller size of CMS, a larger proportion of the decays will occur outside the detector for a specific BP as compared to the ATLAS detector, and therefore for small enough $c \tau$, CMS limits become better than those of ATLAS.
This demonstrates an important complementarity of two detectors.
 Furthermore, the effect of the LL particle mass is visible, because this effects the relativistic $\gamma$-factor and therefore its lifetime in the lab-frame. As an example, in Fig.~\ref{fig:FFX_limit}(e,f), cross section limits for each BP for large $c \tau$ are of the same order of magnitude where near 100\% of decays occur outside the detector for each $m_X$. However for smaller $c \tau$ there is a clear pattern of limits for BPs with lower $m_X$ extending further left into the low $c \tau$ region because of the relativistic time-dilation.

\begin{figure}[htpb]
  \centering
  \includegraphics[width=0.8\linewidth]{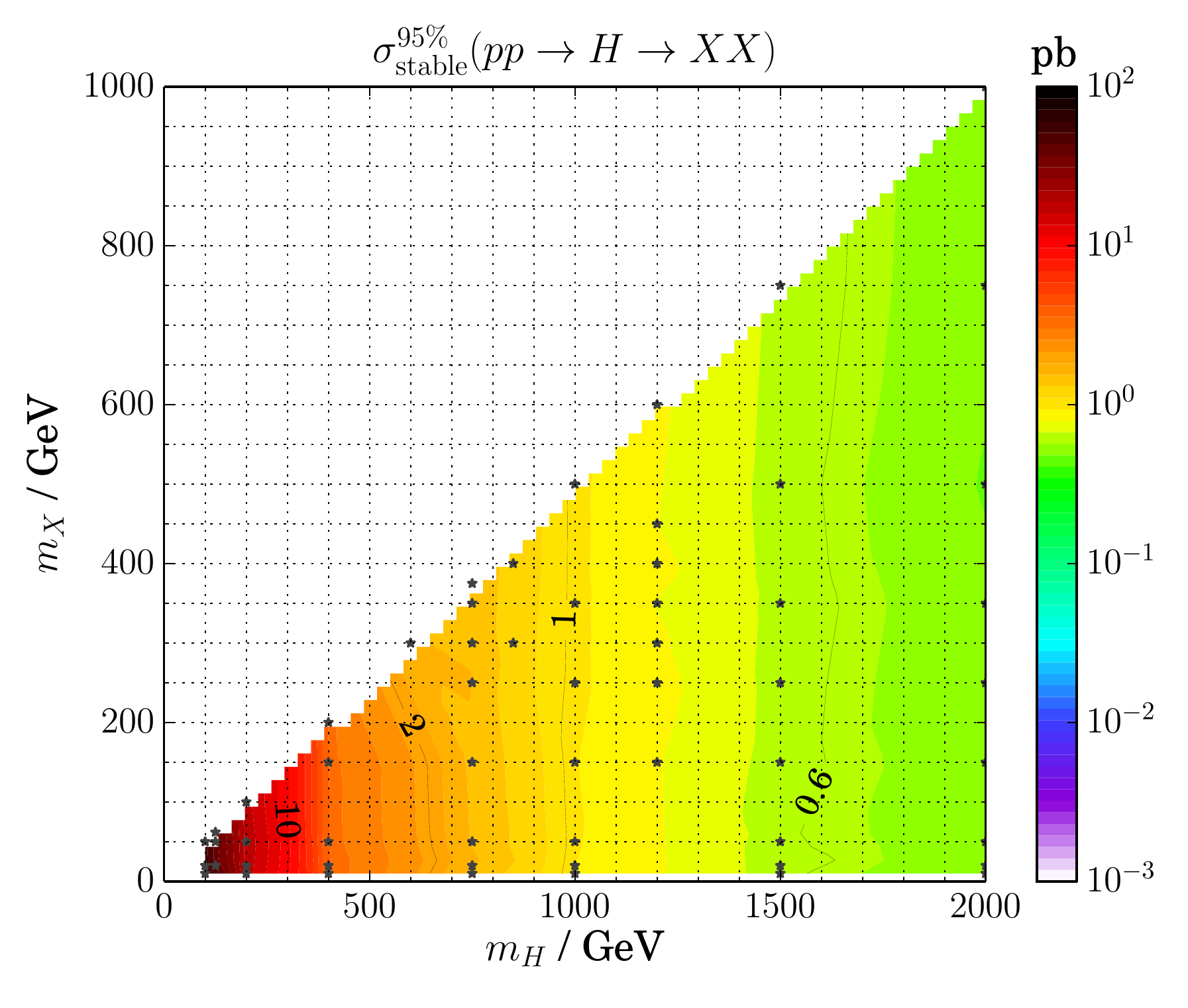}
  \caption{The figure shows the upper limit of the production cross section of $XX$+jets final states for the HXX--model in units of pb. The $x$-axis shows the mass of the mediator $H$ and the $y$-axis shows the mass of the LL $X$ particle. \label{fig:grid-HXX}}
\end{figure}

\begin{figure}[htpb]
  \centering
  \includegraphics[width=0.8\linewidth]{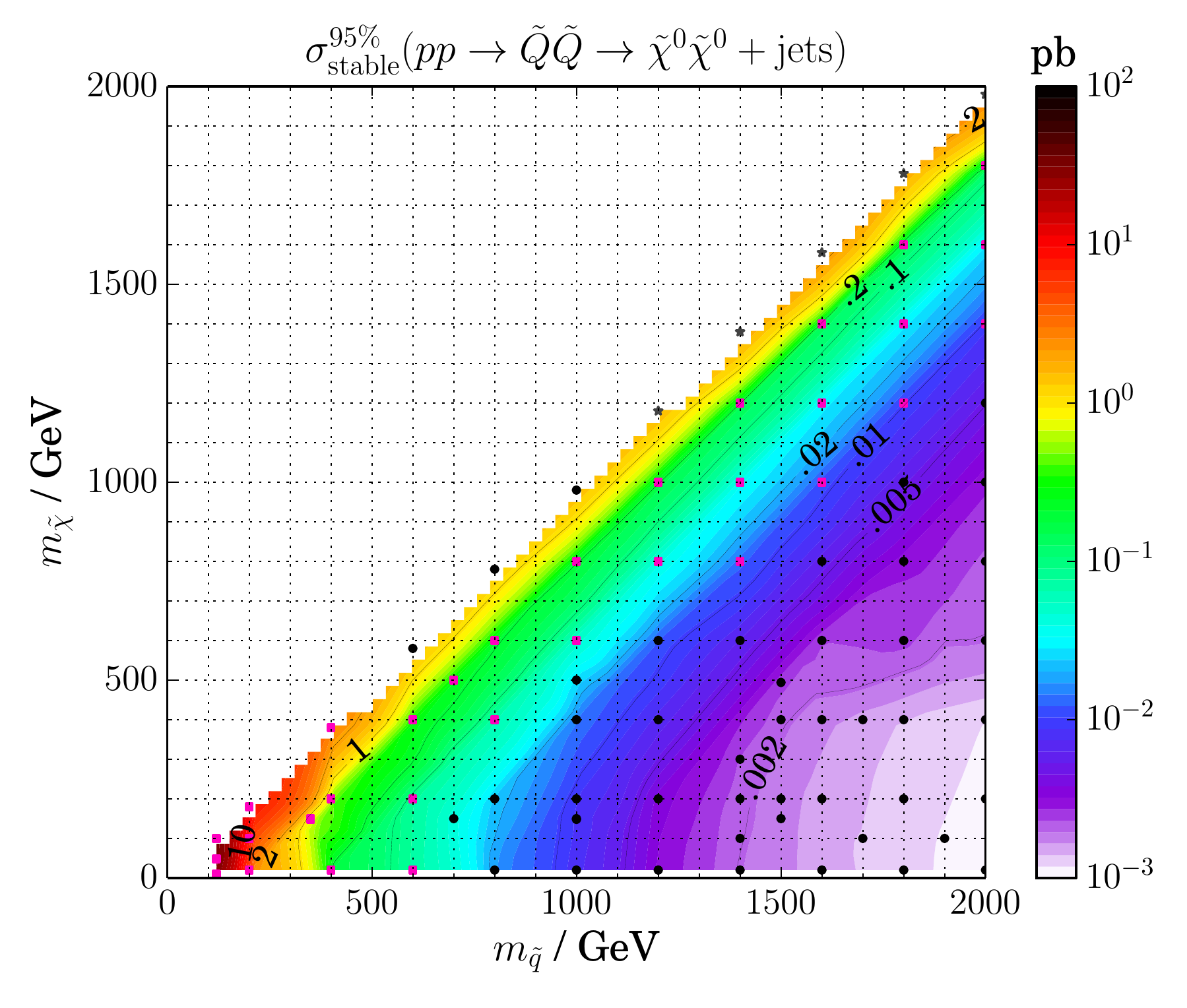}
  \caption{The figure shows the upper limit of the production cross section of $\tilde{\chi}^0\tilde{\chi}^0$+jets final states for the MSSM in units of pb. The $x$-axis shows the mass of the $\tilde q$ squark and the $y$-axis shows the mass of the neutralino $\tilde{\chi}^0$ (LSP). Black dots indicate sample points where the ATLAS multijet paper \cite{TheATLAScollaboration:2013fha} performed best, grey stars indicate the 
ATLAS monojet paper \cite{Aad:2015zva}, and pink squares indicate best performance with the CMS
$\alpha_T$ paper \cite{Chatrchyan:2013mys}. A similar plot showing the different signal regions is shown in the appendix, Fig.~\ref{fig:susy-grid-labels}. \label{fig:grid-susy}}
\end{figure}

To provide a more comprehensive result than only the BPs for these two models, we also present our results for $\sigmastable$ in the form of a $\mmed$ vs $\mLL$ plane. These are shown in Fig.~\ref{fig:grid-HXX} for the HXX model and Fig.~\ref{fig:grid-susy} for the \RpV-SUSY model, where the $\sigmastable$ upper limits are indicated by the colour chart. In order to avoid large statistical errors, for some points, most notably those with large $m_{\tilde{q}}$, we were required to generate very large numbers of Monte Carlo events. It is interesting to note the different pattern of results observed between the HXX model in Fig.~\ref{fig:grid-HXX}, where the cross section limits depend almost exclusively on the mass of the heavy Higgs ($m_H$) mediating the LL $X$ production and the \RpV-SUSY model in Fig.~\ref{fig:grid-susy}, where the limits depend largely on the mass gap $\Delta m = m_{\tilde{q}} - m_{\tilde{\chi}^0}$. 

These differences can be explained by the production and decay channels of the two models. In the HXX model, the heavy Higgs, $H$ is produced on-shell, before decaying into two $X$ bosons, and therefore, the $\ETmis$ is just the $p_T$ of the $H$. As $m_H$ increases, so does its average $p_T$, and therefore so does the $\ETmis$ on which the analysis depends, leading to more stringent cross section bounds with larger $m_H$.

In the $\RpV$-SUSY model on the other hand, each squark decays into a neutralino and quark ($\tilde{q} \to q \tilde{\chi}^0$), giving a signal of $\ETmis$ and jets. For small mass gaps between the squark and the neutralino, $\Delta m$, the decay products tend to be soft, giving a low $\ETmis$, soft jets and a low signal efficiency. In this case, the best limits come from the monojet analysis (or $\alpha_T$ analyses with 
2-jet signatures for low values of $m_{\tilde q}$) and are of similar size as for the Higgs boson model (Fig.~\ref{fig:susy-grid-labels} in the appendix shows the best analysis sets of the ATLAS monojet analysis for each sample point).
As $\Delta m$ increases, the $\ETmis$ and jet $p_T$ increase too, increasing the signal cut efficiency and improving the cross section limits. Therefore the most important parameter for the \RpV-SUSY model is $\Delta m = m_{\tilde q} - m_{\tilde{\chi}}$ as is clearly seen in Fig.~\ref{fig:grid-susy}.

At this point it is worth noting that, whilst we have been working with the HXX model (described in Sec.~\ref{sec:models}) which specifically has a scalar LL particle $X$, these limits are valid regardless of the spin or decay pattern of $X$ and are in fact valid for any model where the production of a scalar, $H$, with a narrow width decaying to a LL particle can be described by the effective vertex $\tfrac{1}{2}\mathrm{Tr}[G^2]H$. This is because the $H$ is a scalar which is produced on-shell, then decays to $X$ pairs isotropically in its rest frame, which then leaves the detector before decaying (for our $\ETmis$ signal), meaning that the spin of $X$ is not relevant.

We should also stress that as we did not simulate displaced vertices explicitly, we do not consider events where one of the particles decay within, and the other outside detector and as a result our limits are conservative. This could potentially give an additional $\ETmis$ signature, particularly for the HXX model where the signal would no longer be suppressed by the requirement of recoiling against a high-$p_T$ jet. Simulations, in particular detector simulations, involving displaced vertices are more technically difficult and therefore was beyond the scope of this study. However as this scenario has the potential to produce strong limits, the authors would like to encourage experimentalists to consider such scenarios.

\section{Conclusion}
\label{sec:conclusion}
CMS and ATLAS have historically searched for long-lived, neutral particles by looking for evidence of their 
decay products within the detector. We have demonstrated  that using missing transverse energy ($\ETmis$) analyses, which are 
traditionally used for dark matter searches, it is possible to complement these existing LHC searches, 
extending the cross section limits on long-lived, neutral particles to arbitrarily long lifetimes.
We have illustrated this by using $\ETmis$ signatures to place cross section limits on two signal models considered in 
CMS searches for displaced leptons or jets produced by long-lived particle decay \cite{CMS:2014hka, CMS:2014wda}.
The limits we obtained using $\ETmis$ are comparable to those from the displaced lepton/jet searches for $c \tau$ values as short as order of a few metres (lifetime of order of a nanosecond), although for the majority of benchmark points, they become comparable at larger distances of $\mathcal{O}$(10m - 100m)
which is the order of the detector size and larger. However it is important to note that whilst our limits on the production cross sections of the LL particles are independent of how the particle decays, the CMS displaced vertex limits depend on the branching ratios to the channels considered. For realistic branching ratios of order a few percent or less, the CMS limits from displaced vertex searches would be considerably weakened and limits from $\ETmis$ become much more competitive. In this case our cross section limits for stable particles can be better then the minimum obtained (for any $c \tau$) from displaced vertex searches, and our new limits can be comparable to those from the displaced vertex searches for decay distances less than 1 metre. 

In the case of a model where a heavy Higgs boson decays to a long-lived, neutral scalar, we used predominantly an ATLAS study of events with a 
monojet and large $\ETmis$ to establish the best limit on the inclusive cross section $\sigma(pp\to H^0\to XX)$. As the signal in this case is suppressed by the requirement of high-$P_T$ ISR jet, the limits are generally weak when compared to the \RpV SUSY model and the 95\% CL cross section limits are of the order of 1 pb or above under the assumption that the particle is stable. Extending  this limit for finite lifetimes (the lifetime given in terms of $c \tau$), depending on the benchmark point,  we found that our new results  improve the published CMS limits for $c \tau$ above  few metres in 
the best cases and for  $c \tau$ above a kilometre in some worst case scenario, corresponding to lifetimes in the nanosecond to microsecond range. 
Furthermore, whilst the CMS displaced lepton/jet papers assume that the long-lived scalar particle  $X$ has specific decay modes, our analysis 
and the limit it gives are valid for any decay mode. This is an additional advantage of using the $\ETmis$ signatures to search for long-lived,
neutral particles.

In the case of an \RpV SUSY model, in which long-lived neutralinos are produced via squark decay, the 95\% CL cross section limits obtained 
for stable particles, using the $\ETmis$ signal, are stronger than the corresponding limits obtained for the Higgs boson model,
and can be as good as approximately 10 fb in case of a large mass splitting between the neutralino and squark. 
In this case the best limits generally come from ATLAS and CMS papers on multijet events in association with large $\ETmis$. 
Also in this case, we then reinterpreted these results to produced upper limits for  the inclusive cross section for $pp\to\tilde Q\tilde Q\to \tilde{\chi}^0\tilde{\chi}^0+\text{jets}$ process as a function of the neutralino lifetime.
  Whilst we derived our limits assuming a specific \RpV-SUSY model, as our limits do not depend on the decay channel of the neutralinos, these limits are valid for any SUSY model with the same production channel 
  assuming negligible effect from heavy intermediate gluino exchange.

We summarise our results in two plots in the form of a $\mmed$ vs $\mLL$ plane in Figs.~\ref{fig:grid-HXX} and \ref{fig:grid-susy}. 
These plots are complemented with the respective Tabs.~\ref{tab:grid-table-susy} and \ref{tab:grid-table-HXX}
presented in the appendix. These tables  containing limits for the grid 
in the  $\mmed$ vs $m_{LL}$ plane could be used  for the interpretation
of various new physics models obtaining the
 dependence on $c \tau$ by a similar procedure as described in Sec.~\ref{sec:escapeprob}.
Similar methods are planned to be used and the respective analysis are planned for LHC run II.

\section*{Acknowledgements}
AB, SM and MCT acknowledge partial support from the STFC grant number  ST/L000296/1
and the NExT Institute. 
AB thanks the Royal Society Leverhulme Trust Senior Research Fellowship LT140094.
MCT acknowledges support from an STFC STEP award. 
KN thanks SM for the hospitality and support he received at the U. of Southampton. He also acknowledges support from BMBF 00160287 \textit{Comparing LHC Data with Beyond the Standard Model Physics}.
AB and MCT also acknowledge partial funding by a Soton-FAPESP grant, and 
ICTP South American Institute for Fundamental Research (ICTP-SAIFR)  in S\~{a}o Paulo support
at the completion stage of the paper.
\begin{appendix}
\allowdisplaybreaks

\section{Escape probability approximation}
\label{sec:escapeprobApp}

In Sec.~\ref{sec:escapeprob} we described a method to extrapolate $\sigmactau$ from the cross section limit for stable particles, $\sigmastable$, which is based on MC events only. There is however a simpler method to get the lifetime dependence, assuming that the LL particles are produced isotropically and their energies are not correlated. While these assumptions are not necessarily accurate, we found that the results are fairly comparable to the ones from the pure-MC method. Other researchers are able to produce similar limits as those in Figs.~\ref{fig:FFX_limit}-\ref{fig:RPV_limit}, given that they know the energy distribution of the LL particles, $g_m(E)$. 

\begin{figure}[h]
  \centering
 \makebox[\textwidth][c]{
  \includegraphics[width=0.54\linewidth]{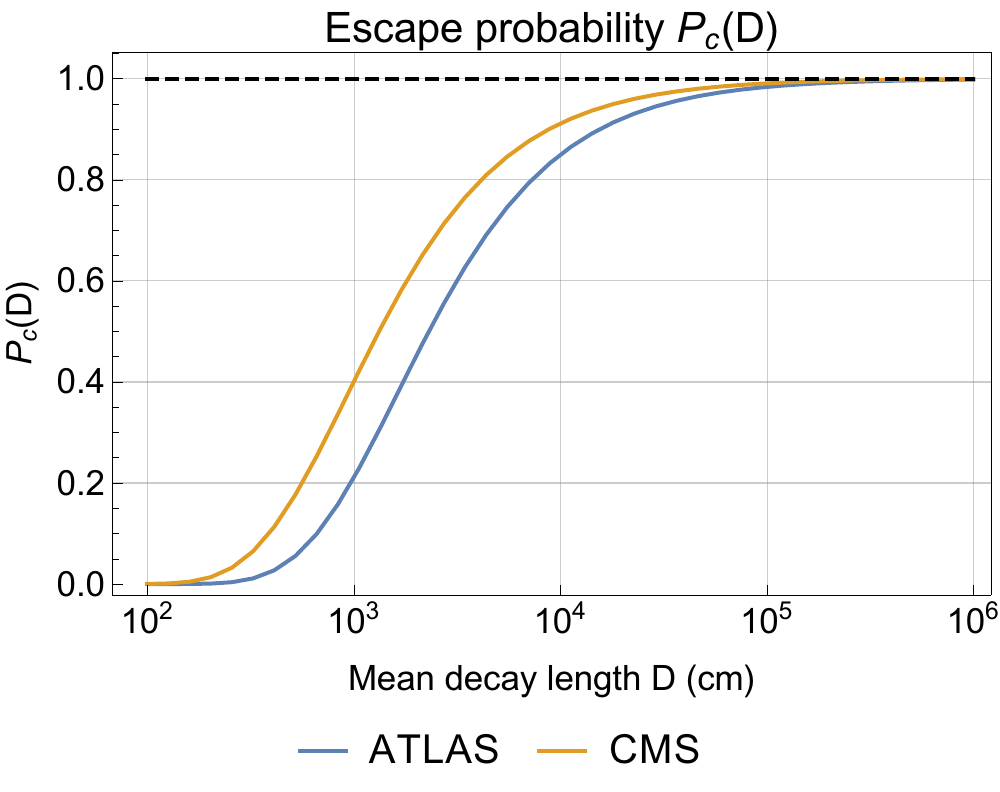}
  \hspace{0mm}
  \includegraphics[width=0.54\linewidth]{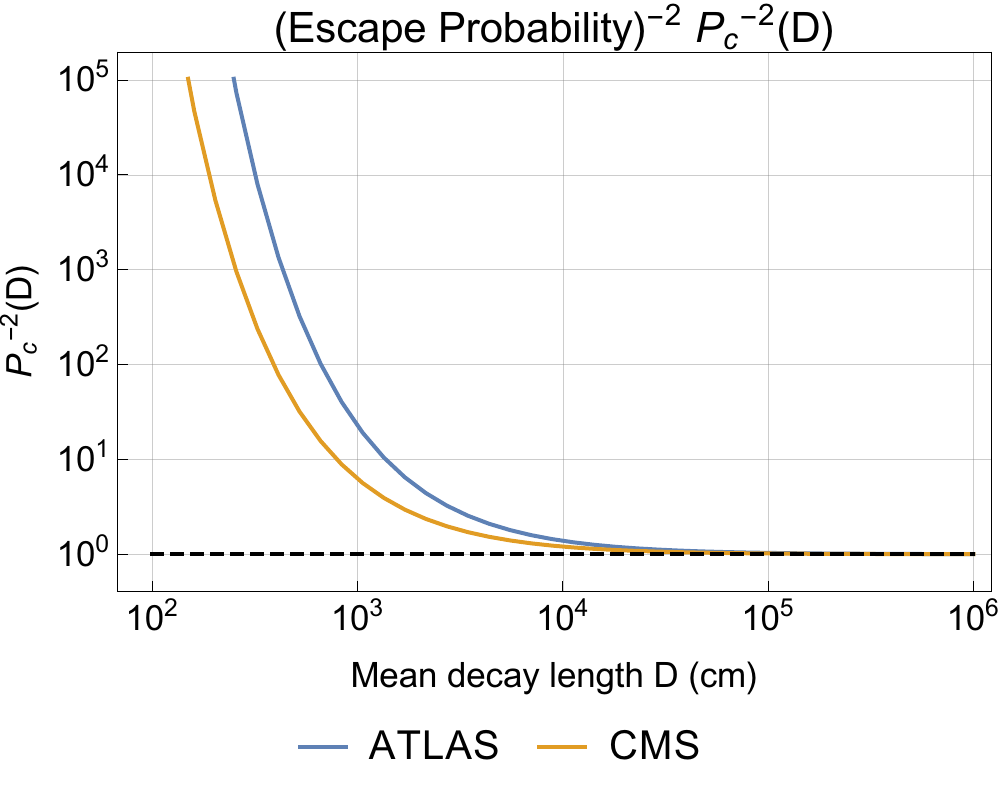}}
  \hspace*{11mm}(a)\hspace*{0.53\textwidth}(b)
  \caption{Escape probability $P_{\text{c}}(D = \beta\gamma c\tau)$ of long lived particles within a detector (left). On the right side,
    $P_{\text{c}}^{-2}(D = \gamma c\tau)$ is shown.}
  \label{fig:prob}
\end{figure}

Defining the distance that a LL particle travelled when it decays, $r = c \beta\gamma t$, and the mean decay distance, $D = c \beta\gamma \tau$, we have that the probability of decaying {\it beyond} a certain distance $r$ is $\exp(-r/D)$. When we also take into account the $1/(4 \pi r^2)$ drop due to increasing area, the probability of a particle crossing a small area $S$ at a boundary at a distance $r$ from the origin is
\begin{align}
P(D) = \int_{S} f(r,D) \de S
\end{align}
with
\begin{align}
f(r,D) &=  \frac{1}{4\pi r^2}\exp \left(\frac{-r}{D}\right).
\end{align}
In our case we wish to find the probability of a particle reaching beyond the boundary of a cylindrical detector, $P_{\text{c}}(D)$. Splitting the cylinder into a barrel and endcaps, we have, 
\begin{align}
  \label{eq:probability}
  P_{\text{c}}(D) &= \int_{\text{barre}l}f(r,D) \de S + \int_{\text{endcap}}f(r,D) \de S \\
            &= 4\pi R \int_0^{L/2} f(\sqrt{z^2+R^2}, D)\frac {R}{\sqrt{z^2+R^2 }}\de z \\
            &+ 4 \pi \int_{0}^{R}f(\sqrt{(L/2)^2+\rho^2}, D)\frac{L/2}{\sqrt{(L/2)^2+\rho^2}}\rho\de \rho. \label{eq:prob} 
\end{align}
The function $P_{\text{c}}(D)$ is universal and its numerical evaluation is shown in Fig.~\ref{fig:prob}. To obtain a probability as a function of $c\tau$, we need to integrate over the relativistic factors $\beta\gamma=\sqrt{\gamma^2-1}$, or substituting $\gamma=E/m$, equivalently integrate over energy $E$. The integration has to be weighted with an energy distribution function, $g_m(E)$, which can be extracted from Monte Carlo simulation of events for each model and mass considered.
The resulting function is $\bar P_{\text{c}}(c\tau)$, which is also mass dependent because of the substitution $\gamma=E/m$.
\begin{align}
  \bar P_{\text{c}}(c\tau) &= \int \de E\, g_m(E)\, P_{\text{c}}(D).
\label{eq:E_int}
\end{align}
Since only the fraction of $\bar P_{\text{c}}(c\tau)$ will contribute to the $\ETmis$ signature in case of unstable particles, we have the relation
\begin{align}
\sigmactau \times \bar P_{\text{c}}(c\tau)^2 = \sigmastable,
\end{align}
which can be rearranged into 
\begin{align}
\label{eq:sigmaformula}\sigmactau = \sigmastable \times \lbrack \bar P_{\text{c}}(c\tau) \rbrack^{-2}.
\end{align}
This equation (\ref{eq:sigmaformula}) provides the needed relation between $\sigmactau$ and $\sigmastable$. 

\section{Tables}
\label{sec:tables}

\begin{center}
  \centering
  \begin{longtable}{|c|c|c|c|c|}
  \caption[Complete result, SUSY grid]{The complete grid scan result of the \RpV-SUSY model. }
  \label{tab:grid-table-susy} \\
    \hline
    $m_{\tilde q}$ (GeV) & $m_{\tilde\chi}$ (GeV) &  analysis & analysis set & $\sigmastable$ (pb) \\ \hline \hline
    \endfirsthead
    %
    \hline
    $m_{\tilde q}$ (GeV) & $m_{\tilde\chi}$ (GeV) &  analysis & analysis set & $\sigmastable$ (pb) \\ \hline \hline
    \endhead
    %
    \endfoot
    %
    \endlastfoot
120 & 10 & CMS $\alpha_T$ \cite{Chatrchyan:2013mys} & 23j\_0b\_275 & 2.96e+01 \\ \hline
120 & 48 & CMS $\alpha_T$ \cite{Chatrchyan:2013mys} & 4j\_0b\_325 & 3.35e+01 \\ \hline
120 & 100 & CMS $\alpha_T$ \cite{Chatrchyan:2013mys} & 23j\_0b\_375 & 3.36e+01 \\ \hline
200 & 20 & CMS $\alpha_T$ \cite{Chatrchyan:2013mys} & 23j\_0b\_275 & 2.46e+00 \\ \hline
200 & 100 & CMS $\alpha_T$ \cite{Chatrchyan:2013mys} & 4j\_0b\_325 & 5.00e+00 \\ \hline
200 & 180 & CMS $\alpha_T$ \cite{Chatrchyan:2013mys} & 23j\_0b\_325 & 8.77e+00 \\ \hline
350 & 148 & CMS $\alpha_T$ \cite{Chatrchyan:2013mys} & 23j\_0b\_325 & 5.73e-01 \\ \hline
350 & 150 & CMS $\alpha_T$ \cite{Chatrchyan:2013mys} & 23j\_0b\_325 & 5.33e-01 \\ \hline
400 & 20 & CMS $\alpha_T$ \cite{Chatrchyan:2013mys} & 23j\_0b\_375 & 1.71e-01 \\ \hline
400 & 200 & CMS $\alpha_T$ \cite{Chatrchyan:2013mys} & 23j\_0b\_375 & 4.27e-01 \\ \hline
400 & 380 & CMS $\alpha_T$ \cite{Chatrchyan:2013mys} & 23j\_0b\_375 & 2.68e+00 \\ \hline
600 & 20 & CMS $\alpha_T$ \cite{Chatrchyan:2013mys} & 23j\_0b\_675 & 7.33e-02 \\ \hline
600 & 200 & CMS $\alpha_T$ \cite{Chatrchyan:2013mys} & 23j\_0b\_475 & 7.74e-02 \\ \hline
600 & 400 & CMS $\alpha_T$ \cite{Chatrchyan:2013mys} & 23j\_0b\_375 & 2.75e-01 \\ \hline
600 & 580 & ATLAS multijet \cite{TheATLAScollaboration:2013fha} & AM & 1.56e+00 \\ \hline
700 & 150 & ATLAS multijet \cite{TheATLAScollaboration:2013fha} & AM & 4.14e-02 \\ \hline
700 & 500 & CMS $\alpha_T$ \cite{Chatrchyan:2013mys} & 23j\_0b\_375 & 2.43e-01 \\ \hline
800 & 20 & ATLAS multijet \cite{TheATLAScollaboration:2013fha} & AM & 1.62e-02 \\ \hline
800 & 200 & ATLAS multijet \cite{TheATLAScollaboration:2013fha} & AM & 2.24e-02 \\ \hline
800 & 400 & CMS $\alpha_T$ \cite{Chatrchyan:2013mys} & 23j\_0b\_675 & 6.69e-02 \\ \hline
800 & 600 & CMS $\alpha_T$ \cite{Chatrchyan:2013mys} & 23j\_0b\_375 & 2.15e-01 \\ \hline
800 & 780 & ATLAS multijet \cite{TheATLAScollaboration:2013fha} & AM & 1.39e+00 \\ \hline
1000 & 20 & ATLAS multijet \cite{TheATLAScollaboration:2013fha} & AM & 7.80e-03 \\ \hline
1000 & 148 & ATLAS multijet \cite{TheATLAScollaboration:2013fha} & AM & 8.56e-03 \\ \hline
1000 & 150 & ATLAS multijet \cite{TheATLAScollaboration:2013fha} & AM & 8.38e-03 \\ \hline
1000 & 200 & ATLAS multijet \cite{TheATLAScollaboration:2013fha} & AM & 8.68e-03 \\ \hline
1000 & 400 & ATLAS multijet \cite{TheATLAScollaboration:2013fha} & AM & 1.40e-02 \\ \hline
1000 & 500 & ATLAS multijet \cite{TheATLAScollaboration:2013fha} & AM & 2.54e-02 \\ \hline
1000 & 600 & CMS $\alpha_T$ \cite{Chatrchyan:2013mys} & 23j\_0b\_675 & 4.80e-02 \\ \hline
1000 & 800 & CMS $\alpha_T$ \cite{Chatrchyan:2013mys} & 23j\_0b\_375 & 1.87e-01 \\ \hline
1000 & 980 & ATLAS multijet \cite{TheATLAScollaboration:2013fha} & AM & 1.55e+00 \\ \hline
1200 & 20 & ATLAS multijet \cite{TheATLAScollaboration:2013fha} & CT & 3.05e-03 \\ \hline
1200 & 200 & ATLAS multijet \cite{TheATLAScollaboration:2013fha} & CT & 3.49e-03 \\ \hline
1200 & 400 & ATLAS multijet \cite{TheATLAScollaboration:2013fha} & AM & 6.89e-03 \\ \hline
1200 & 600 & ATLAS multijet \cite{TheATLAScollaboration:2013fha} & AM & 1.14e-02 \\ \hline
1200 & 800 & CMS $\alpha_T$ \cite{Chatrchyan:2013mys} & 23j\_0b\_675 & 3.90e-02 \\ \hline
1200 & 1000 & CMS $\alpha_T$ \cite{Chatrchyan:2013mys} & 23j\_0b\_375 & 1.60e-01 \\ \hline
1200 & 1180 & ATLAS monojet \cite{Aad:2015zva} & SR7 & 1.57e+00 \\ \hline
1400 & 20 & ATLAS multijet \cite{TheATLAScollaboration:2013fha} & CT & 1.96e-03 \\ \hline
1400 & 100 & ATLAS multijet \cite{TheATLAScollaboration:2013fha} & CT & 2.02e-03 \\ \hline
1400 & 200 & ATLAS multijet \cite{TheATLAScollaboration:2013fha} & CT & 2.07e-03 \\ \hline
1400 & 300 & ATLAS multijet \cite{TheATLAScollaboration:2013fha} & CT & 2.25e-03 \\ \hline
1400 & 600 & ATLAS multijet \cite{TheATLAScollaboration:2013fha} & AM & 6.11e-03 \\ \hline
1400 & 800 & CMS $\alpha_T$ \cite{Chatrchyan:2013mys} & 23j\_0b\_875 & 1.29e-02 \\ \hline
1400 & 1000 & CMS $\alpha_T$ \cite{Chatrchyan:2013mys} & 23j\_0b\_675 & 3.44e-02 \\ \hline
1400 & 1200 & CMS $\alpha_T$ \cite{Chatrchyan:2013mys} & 23j\_0b\_375 & 1.48e-01 \\ \hline
1400 & 1380 & ATLAS monojet \cite{Aad:2015zva} & SR8 & 1.94e+00 \\ \hline
1500 & 150 & ATLAS multijet \cite{TheATLAScollaboration:2013fha} & CT & 1.79e-03 \\ \hline
1500 & 200 & ATLAS multijet \cite{TheATLAScollaboration:2013fha} & CT & 1.76e-03 \\ \hline
1500 & 400 & ATLAS multijet \cite{TheATLAScollaboration:2013fha} & CT & 2.06e-03 \\ \hline
1500 & 494 & ATLAS multijet \cite{TheATLAScollaboration:2013fha} & CT & 2.40e-03 \\ \hline
1600 & 20 & ATLAS multijet \cite{TheATLAScollaboration:2013fha} & CT & 1.51e-03 \\ \hline
1600 & 200 & ATLAS multijet \cite{TheATLAScollaboration:2013fha} & CT & 1.53e-03 \\ \hline
1600 & 400 & ATLAS multijet \cite{TheATLAScollaboration:2013fha} & CT & 1.73e-03 \\ \hline
1600 & 600 & ATLAS multijet \cite{TheATLAScollaboration:2013fha} & CT & 2.31e-03 \\ \hline
1600 & 800 & ATLAS multijet \cite{TheATLAScollaboration:2013fha} & AM & 5.68e-03 \\ \hline
1600 & 1000 & CMS $\alpha_T$ \cite{Chatrchyan:2013mys} & 23j\_0b\_875 & 1.11e-02 \\ \hline
1600 & 1200 & CMS $\alpha_T$ \cite{Chatrchyan:2013mys} & 23j\_0b\_675 & 3.06e-02 \\ \hline
1600 & 1400 & CMS $\alpha_T$ \cite{Chatrchyan:2013mys} & 23j\_0b\_375 & 1.38e-01 \\ \hline
1600 & 1580 & ATLAS monojet \cite{Aad:2015zva} & SR8 & 2.34e+00 \\ \hline
1700 & 100 & ATLAS multijet \cite{TheATLAScollaboration:2013fha} & CT & 1.35e-03 \\ \hline
1700 & 400 & ATLAS multijet \cite{TheATLAScollaboration:2013fha} & CT & 1.54e-03 \\ \hline
1800 & 20 & ATLAS multijet \cite{TheATLAScollaboration:2013fha} & CT & 1.23e-03 \\ \hline
1800 & 200 & ATLAS multijet \cite{TheATLAScollaboration:2013fha} & CT & 1.27e-03 \\ \hline
1800 & 400 & ATLAS multijet \cite{TheATLAScollaboration:2013fha} & CT & 1.38e-03 \\ \hline
1800 & 600 & ATLAS multijet \cite{TheATLAScollaboration:2013fha} & BT & 2.43e-03 \\ \hline
1800 & 800 & ATLAS multijet \cite{TheATLAScollaboration:2013fha} & BT & 2.97e-03 \\ \hline
1800 & 1000 & ATLAS multijet \cite{TheATLAScollaboration:2013fha} & AM & 5.39e-03 \\ \hline
1800 & 1200 & CMS $\alpha_T$ \cite{Chatrchyan:2013mys} & 23j\_0b\_875 & 9.88e-03 \\ \hline
1800 & 1400 & CMS $\alpha_T$ \cite{Chatrchyan:2013mys} & 23j\_0b\_675 & 2.84e-02 \\ \hline
1800 & 1600 & CMS $\alpha_T$ \cite{Chatrchyan:2013mys} & 23j\_0b\_375 & 1.26e-01 \\ \hline
1800 & 1780 & ATLAS monojet \cite{Aad:2015zva} & SR6 & 1.80e+00 \\ \hline
1900 & 100 & ATLAS multijet \cite{TheATLAScollaboration:2013fha} & CT & 1.17e-03 \\ \hline
2000 & 20 & ATLAS multijet \cite{TheATLAScollaboration:2013fha} & CT & 1.10e-03 \\ \hline
2000 & 200 & ATLAS multijet \cite{TheATLAScollaboration:2013fha} & CT & 1.13e-03 \\ \hline
2000 & 400 & ATLAS multijet \cite{TheATLAScollaboration:2013fha} & CT & 1.18e-03 \\ \hline
2000 & 600 & ATLAS multijet \cite{TheATLAScollaboration:2013fha} & BT & 1.99e-03 \\ \hline
2000 & 800 & ATLAS multijet \cite{TheATLAScollaboration:2013fha} & BT & 2.21e-03 \\ \hline
2000 & 1000 & ATLAS multijet \cite{TheATLAScollaboration:2013fha} & BT & 2.83e-03 \\ \hline
2000 & 1200 & ATLAS multijet \cite{TheATLAScollaboration:2013fha} & BT & 5.02e-03 \\ \hline
2000 & 1400 & CMS $\alpha_T$ \cite{Chatrchyan:2013mys} & 23j\_0b\_875 & 9.19e-03 \\ \hline
2000 & 1600 & CMS $\alpha_T$ \cite{Chatrchyan:2013mys} & 23j\_0b\_675 & 2.65e-02 \\ \hline
2000 & 1800 & CMS $\alpha_T$ \cite{Chatrchyan:2013mys} & 23j\_0b\_375 & 1.21e-01 \\ \hline
2000 & 1980 & ATLAS monojet \cite{Aad:2015zva} & SR7 & 2.71e+00 \\ \hline

  \end{longtable}
\end{center}

\begin{center}
  \centering
  \begin{longtable}{|c|c|c|c|c|}
  \caption[Complete result, HXX grid]{The complete grid scan result of the HXX model. }
  \label{tab:grid-table-HXX} \\
    \hline
    $m_{H}$ (GeV) & $m_{X}$ (GeV) &  analysis & analysis set & $\sigmastable$ (pb) \\ \hline \hline
    \endfirsthead
    %
    \hline
    $m_{H}$ (GeV) & $m_{X}$ (GeV) &  analysis & analysis set & $\sigmastable$ (pb) \\ \hline \hline
    \endhead
    %
    \endfoot
    %
    \endlastfoot
100 & 10 & ATLAS monojet \cite{Aad:2015zva} & SR4 & 5.77e+01\\ \hline
100 & 20 & ATLAS monojet \cite{Aad:2015zva} & SR4 & 5.58e+01\\ \hline
100 & 50 & ATLAS monojet \cite{Aad:2015zva} & SR4 & 5.38e+01\\ \hline
125 & 20 & ATLAS monojet \cite{Aad:2015zva} & SR4 & 3.83e+01\\ \hline
125 & 50 & ATLAS monojet \cite{Aad:2015zva} & SR4 & 3.99e+01\\ \hline
125 & 62 & ATLAS monojet \cite{Aad:2015zva} & SR4 & 3.79e+01\\ \hline
200 & 10 & ATLAS monojet \cite{Aad:2015zva} & SR4 & 1.67e+01\\ \hline
200 & 20 & ATLAS monojet \cite{Aad:2015zva} & SR4 & 1.71e+01\\ \hline
200 & 50 & ATLAS monojet \cite{Aad:2015zva} & SR4 & 1.75e+01\\ \hline
200 & 100 & ATLAS monojet \cite{Aad:2015zva} & SR4 & 1.65e+01 \\ \hline
400 & 10 & ATLAS monojet \cite{Aad:2015zva} & SR6 & 3.26e+00\\ \hline
400 & 20 & ATLAS monojet \cite{Aad:2015zva} & SR6 & 3.29e+00\\ \hline
400 & 50 & ATLAS monojet \cite{Aad:2015zva} & SR6 & 3.17e+00\\ \hline
400 & 150 & ATLAS monojet \cite{Aad:2015zva} & SR6 & 3.16e+00\\ \hline
400 & 200 & ATLAS monojet \cite{Aad:2015zva} & SR6 & 3.12e+00\\ \hline
600 & 300 & ATLAS monojet \cite{Aad:2015zva} & SR6 & 1.57e+00\\ \hline
750 & 10 & ATLAS monojet \cite{Aad:2015zva} & SR7 & 1.48e+00\\ \hline
750 & 20 & ATLAS monojet \cite{Aad:2015zva} & SR7 & 1.61e+00\\ \hline
750 & 50 & ATLAS monojet \cite{Aad:2015zva} & SR7 & 1.51e+00\\ \hline
750 & 150 & ATLAS monojet \cite{Aad:2015zva} & SR7 & 1.48e+00\\ \hline
750 & 250 & ATLAS monojet \cite{Aad:2015zva} & SR7 & 1.57e+00\\ \hline
750 & 300 & ATLAS monojet \cite{Aad:2015zva} & SR7 & 1.49e+00\\ \hline
750 & 350 & ATLAS monojet \cite{Aad:2015zva} & SR7 & 1.46e+00\\ \hline
750 & 375 & ATLAS monojet \cite{Aad:2015zva} & SR7 & 1.46e+00\\ \hline
850 & 300 & ATLAS monojet \cite{Aad:2015zva} & SR7 & 1.19e+00\\ \hline
850 & 400 & ATLAS monojet \cite{Aad:2015zva} & SR7 & 1.22e+00\\ \hline
1000 & 10 & ATLAS monojet \cite{Aad:2015zva} & SR7 & 9.31e-01\\ \hline
1000 & 20 & ATLAS monojet \cite{Aad:2015zva} & SR7 & 9.39e-01\\ \hline
1000 & 50 & ATLAS monojet \cite{Aad:2015zva} & SR7 & 9.55e-01\\ \hline
1000 & 150 & ATLAS monojet \cite{Aad:2015zva} & SR7 & 9.41e-01\\ \hline
1000 & 250 & ATLAS monojet \cite{Aad:2015zva} & SR7 & 9.64e-01\\ \hline
1000 & 350 & ATLAS monojet \cite{Aad:2015zva} & SR7 & 9.67e-01\\ \hline
1000 & 500 & ATLAS monojet \cite{Aad:2015zva} & SR7 & 9.69e-01\\ \hline
1200 & 150 & ATLAS monojet \cite{Aad:2015zva} & SR8 & 8.00e-01\\ \hline
1200 & 250 & ATLAS monojet \cite{Aad:2015zva} & SR8 & 8.32e-01\\ \hline
1200 & 300 & ATLAS monojet \cite{Aad:2015zva} & SR8 & 8.11e-01\\ \hline
1200 & 350 & ATLAS monojet \cite{Aad:2015zva} & SR8 & 7.85e-01\\ \hline
1200 & 400 & ATLAS monojet \cite{Aad:2015zva} & SR8 & 8.30e-01\\ \hline
1200 & 450 & ATLAS monojet \cite{Aad:2015zva} & SR8 & 7.93e-01\\ \hline
1200 & 600 & ATLAS monojet \cite{Aad:2015zva} & SR8 & 8.11e-01\\ \hline
1500 & 10 & ATLAS monojet \cite{Aad:2015zva} & SR8 & 6.17e-01\\ \hline
1500 & 20 & ATLAS monojet \cite{Aad:2015zva} & SR8 & 6.33e-01\\ \hline
1500 & 50 & ATLAS monojet \cite{Aad:2015zva} & SR8 & 6.10e-01\\ \hline
1500 & 150 & ATLAS monojet \cite{Aad:2015zva} & SR8 & 6.27e-01\\ \hline
1500 & 250 & ATLAS monojet \cite{Aad:2015zva} & SR8 & 6.32e-01\\ \hline
1500 & 350 & ATLAS monojet \cite{Aad:2015zva} & SR8 & 6.41e-01\\ \hline
1500 & 500 & ATLAS monojet \cite{Aad:2015zva} & SR8 & 6.32e-01\\ \hline
1500 & 750 & ATLAS monojet \cite{Aad:2015zva} & SR8 & 6.47e-01\\ \hline
2000 & 10 & ATLAS monojet \cite{Aad:2015zva} & SR8 & 4.90e-01\\ \hline
2000 & 20 & ATLAS monojet \cite{Aad:2015zva} & SR8 & 4.93e-01\\ \hline
2000 & 50 & ATLAS monojet \cite{Aad:2015zva} & SR8 & 5.11e-01\\ \hline
2000 & 150 & ATLAS monojet \cite{Aad:2015zva} & SR8 & 5.10e-01\\ \hline
2000 & 250 & ATLAS monojet \cite{Aad:2015zva} & SR8 & 4.90e-01\\ \hline
2000 & 350 & ATLAS monojet \cite{Aad:2015zva} & SR8 & 4.99e-01\\ \hline
2000 & 500 & ATLAS monojet \cite{Aad:2015zva} & SR8 & 4.72e-01\\ \hline
2000 & 750 & ATLAS monojet \cite{Aad:2015zva} & SR8 & 5.00e-01\\ \hline
2000 & 1000 & ATLAS monojet \cite{Aad:2015zva} & SR8 & 5.06e-01\\ \hline

  \end{longtable}
\end{center}

\begin{figure}
  \centering
  \includegraphics[width=.8\linewidth]{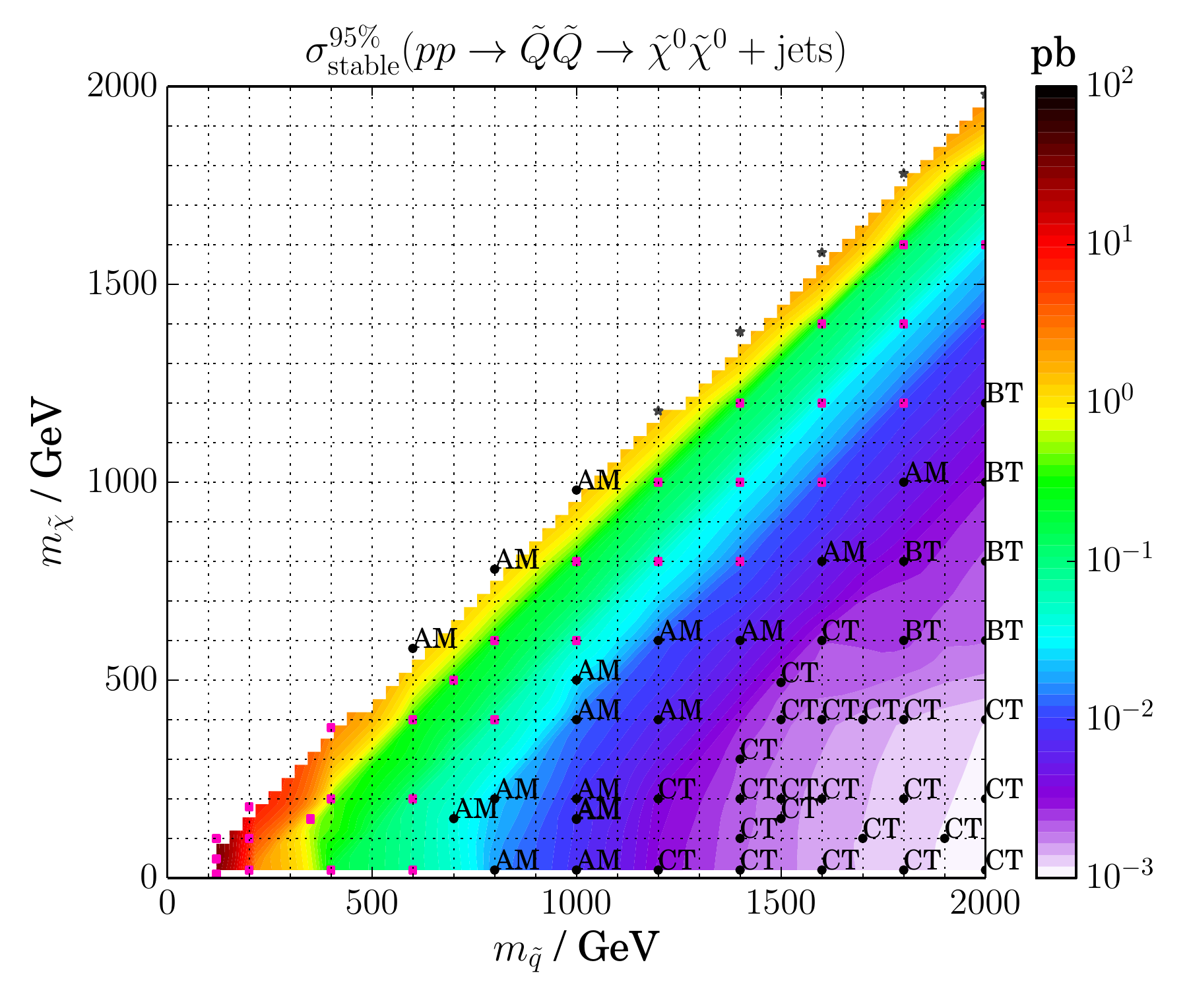}
  \caption{The figure shows the upper limit of the production cross section of $\tilde{\chi}^0\tilde{\chi}^0$+jets final states for the \RpV-MSSM, similar to Fig.~\ref{fig:grid-susy}. The $x$-axis shows the mass of the $\tilde q$ squark and the $y$-axis shows the mass of the neutralino $\tilde{\chi}^0$. 
Black dots indicate sample points where the ATLAS multijet paper \cite{TheATLAScollaboration:2013fha} 
performed best, grey stars indicate the ATLAS monojet paper \cite{Aad:2015zva}, and pink squares indicate 
best performance with the CMS $\alpha_T$ paper \cite{Chatrchyan:2013mys}.
Additionally, the analysis sets (signal regions) for the ATLAS multijet analyses are shown. } 
  \label{fig:susy-grid-labels}
\end{figure}

\end{appendix}

\bibliographystyle{JHEP-2}
\bibliography{literature}

\end{document}